\documentclass{article}

\usepackage[T1]{fontenc}
\usepackage[utf8]{inputenc}
\usepackage{geometry}
\usepackage{fancyhdr}
\usepackage{color}
\usepackage{bm}
\usepackage{amsmath}
\usepackage{amsthm}
\usepackage{amssymb}
\usepackage{graphicx}
\usepackage[english]{babel}
\usepackage{subcaption}
\usepackage{float}
\usepackage{comment}
\usepackage{hyperref}
\usepackage[authoryear]{natbib}
\usepackage{booktabs}
\usepackage{multirow}
\usepackage{siunitx}
\usepackage{makecell}

\makeatletter
\let\ps@plain\ps@fancy   
\makeatother

\usepackage{authblk}
 
\renewcommand{\v}[1]{\boldsymbol{{#1}}} 
\newcommand{\dd}{\delta} 

\title{Three dimensional simulation of fluid-driven frictional and tensile ruptures 
on existing discontinuities
}
\author[]{Brice Lecampion\footnote{Corresponding author, \href{mailto:brice.lecampion@epfl.ch}{brice.lecampion@epfl.ch}}, Sylvain Brisson, Antareep Sarma, Ankit Gupta, Alexis S\'aez, Regina Fakhretdinova}
\affil[]{Geo-Energy Lab - EPFL, Lausanne CH-1015}
\date{May 13, 2026}

\begin{document}

\maketitle

\begin{abstract}
We present an implicit, fully-coupled hydro-mechanical solver for the three-dimensional simulation of fluid-driven rupture propagation along pre-existing discontinuities. The solver handles simultaneously both frictional slip (shear failure) and tensile opening (hydraulic fracture) along arbitrary intersecting fractures and faults in a linearly elastic and impermeable rock matrix. The spatial discretization combines a collocation displacement discontinuity boundary element method for quasi-static elasticity with a Galerkin finite element method for nonlinear pore-fluid diffusion along the discontinuities. Frictional and tensile failure are governed by a poro-elastoplastic cohesive-zone-like interface law with slip-weakening friction, dilatancy, and tensile strength degradation. The interface constitutive relation is integrated via an elastic predictor–plastic corrector scheme. The strong nonlinear coupling between mechanical deformation and fracture permeability changes associated with shear-induced dilation and the mechanical tensile opening of fractures is handled via an adaptive implicit time-stepping scheme. Efficient block preconditioning of the coupled tangent system, leveraging hierarchical matrix representations of the boundary element operator, has proven essential to achieve robustness over a wide range of fracture behaviors (friction, tensile hydraulic failure).
The accuracy and convergence of the solver are demonstrated against a comprehensive suite of analytical and semi-analytical rupture propagation solutions of increasing complexity: self-similar fluid-driven frictional ruptures under constant and slip-weakening friction (including non-circular ruptures and transient dynamic instabilities), dilatant ruptures with permeability changes, and penny-shaped hydraulic fractures spanning the transition between the viscosity and toughness dominated growth regimes. The solver is further assessed on two multi-fracture configurations: injection into three intersecting fractures, and a height-confined hydraulic fracture intersecting a strike-slip fault subsequently slipping. These examples demonstrate that the proposed solver is capable of resolving the coupled hydro-mechanical response of  fracture systems across multiple length and time scales, simultaneously capturing frictional slip, shear-induced dilatancy, permeability evolution, and tensile opening within a unified, verified framework. The solver is well suited for the simulation of fluid-driven ruptures in faulted and fractured subsurface reservoirs.
\end{abstract}

\section{Introduction}

\thispagestyle{empty}
Reactivation of existing discontinuities in the form of fractures and faults in the upper Earth crust due to fluid pressurization - if sustained - leads to the propagation of frictional ruptures along preferably oriented existing discontinuities. These discontinuities can even mechanically open if the fluid pressure exceeds the in-situ normal stress acting on them, resulting in the propagation of tensile/opening hydraulic fracture fronts. Examples range from the effect of large-scale injection on fault movement \citep{HeRu68,HaMe71,Ells13}, interactions of hydraulic fractures with faults \citep{EyEa19} to the stimulation of fractured geothermal reservoirs via hydro-shearing \citep{Jung13,Corn16}. Natural occurrences of fault slip and seismic swarms associated with fluid have also been documented \citep{KaSa10,RoRo17,MaZe24}.

Due to its importance in geo-energy engineering, the hydro-mechanical modeling of fluid-driven ruptures in fractured rocks have received considerable attention over the past fifty years.
Computational challenges abounds for this class of moving boundaries and tightly coupled hydromechanical problems. In particular, fracture permeability changes due to deformation result in very stiff non-linearities. 
Similar to pure tensile hydraulic fracturing, several length and time- scales must be resolved in relation notably to the frictional and possibly opening fronts, permeability changes and fracture intersections. In addition, the potential occurrence of frictional instabilities and the transition to a dynamic rupture pause further computational constraints in terms of spatio-temporal resolution.
Although coupled (thermo)-hydro-mechanical solvers are now well mature in the linear and weakly non-linear cases  (for example without strong permeability changes as in \cite{FeGa08,JhJu14,PrSu16,LiPr19,LiPr19b} among many others), the  
simulation of fracture propagation due to fluid-injection coupled with permeability changes 
remains extremely challenging despite numerous recent contributions that we discuss later. 
Often large scale simulation stops upon fault re-activation, or suffer from a lack of proper spatial resolution to capture fracture growth over large time and length scales \citep{RuRi13,RiRu15}.
 
The development of robust, accurate and computationally efficient solvers have been for a long time impeded due to the lack of available solutions for the propagation of fluid-driven frictional ruptures that could serve as verification tests and further guide numerical developments. 
Similarly than for the case of purely tensile hydraulic fracturing \citep{Deto04,Deto16}, this is no longer the case thanks to analytical and semi-analytical solutions for frictional ruptures in 2 and 3D obtained over the past years \citep{GaGe12,BaVi19,Vies21,SaLe22,SaLe24,Dunh24,Vies25}. 
A series of canonical verification tests now exist for both frictional and tensile fluid-driven fractures that any simulator must critically pass before being used to model more complex configurations - in terms of geometry, frictional rheology, injection sequence and additional physics such as thermal effects. We leverage these analytical fracture propagation solutions to properly argue computational accuracy and efficiency.

A large number of  contributions have been devoted to this class of problems since the early work of \cite{NoAy82,PiCu85}, from a hydraulic fracturing perspective (see e.g.~\cite{LeBu18} for a review), in relation to hydro-mechanical deformation of fractured rocks (see e.g.~\cite{VaYo25} for a review), as well as motivated by  fluid-induced aseismic fault slip \citep{LaEr25}. 
Spatial discretization varies from domain based methods (finite element, finite volume) \citep{FrCa20} to boundary integral equations for the solution of the continuum mechanics problem  \citep{McHo13,KaGh18,CiLe23}. Distinct elements approach either at meso or macro-scale have also been used - from the earlier work of \cite{PiBa84,PiCu85} in two dimensions to 3D cases \citep{CaGu18}. We refer to \cite{DaCu16,DaDe16} for a review of distinct element methods and focus mostly on continuum based models thereafter. 

The proposed numerical solvers can be broadly delineated as function of their choice of time-integration. 
Explicit time-stepping schemes have the advantage of simplicity at the expense of a restrictive condition on the maximum time-step size associated with fluid diffusion: $\Delta t < \mathcal{O}(\Delta x^2)$, with $\Delta x$ the discretization scale. This maximum time-step unfortunately further drops to $\Delta t < \mathcal{O}(\Delta x^3)$ when the  fracture hydromechanically opens in relation to the elasto-lubrication coupling \citep{AdSi07}. 

Explicit time-stepping is popular for fault slip rupture simulations \citep{ChBh25}, and recent works have extended the approach to account for fluid driven aseismic slip. In most cases, the hydraulic properties (permeability) is assumed constant such that the flow and mechanical problems partly uncouple \cite{LaEr25,ImAv24}. Implicit stepping for the simulation of the flow problem is sometimes used in an explicit-implicit strategy \cite{McHo11,RoAm25}. 
Some works  in 2D do account for permeability variation with slip- for example in the framework of rate and state friction \cite{McHo13,NoMc16,OzYa24,BeOz25b}. 
However, hydraulic tensile rupture can not be modeled as mechanical opening is prevented in these frictional-only numerical schemes. 
Explicit time-stepping  is also used in combination with distinct elements and finite volume solver. Examples of combined frictional and opening rupture simulation do exist, but usually without specific demonstration of accuracy against analytical fracture growth solutions with some exceptions \citep{DaDe16}. 

The unconditionally stable property of implicit time-integration schemes (with respect to time-step size) render them particularly suited for these very stiff hydro-mechanical problems. 
In the context of finite element / finite volume methods, the discontinuities either coincides with the mesh (\cite{UcBe18,FrCa20,GaAn26} among others) or modeled via enhanced discretization (X-FEM, extended finite volume) \citep{DeJe17,LiPr19,DeJe20}. 
Boundary elements techniques are also widely used, as they allow high accuracy at reduced cost  \citep{KaGh18,CiLe20,KaGh23} at  the price of uniform elastic properties. 
Frictional contact is solved either via Lagrange multiplier in a rigid-plastic fashion \citep{StBe21,HoPa25,MoPa25}, 
or via the introduction of penalty parameters in an elastoplastic fashion \citep{Wrig06,SaLe22,GaAn26}. On the basis of the well-known non-linear stiffness of rock joints \citep{Corn15}, the later appears more physically justified. 

 Full three dimensional simulations of fluid-driven rupture propagation spanning several length and time-scales are difficult, and clearly lacks a strong suite of verification tests, as exemplified in most of the work previously cited. Better reproducibility should enable the development of more efficient numerical methods for these non-linear multi-physics problems. This contribution is a step in that direction.
We present an implicit time-stepping scheme that is computationally efficient and accurate for the simulation of the propagation of both 1) frictional slip dilatant ruptures  (shear failure) and 2) hydraulic fracture (tensile opening) along existing discontinuities. The two modes of failure (frictional slip and opening) being susceptible to occur simultaneously in a large number of situations, they must be modeled accurately.
The algorithm is directly applicable to multiple, possibly intersecting, fractures in 3D.
We assume from the onset that the rock matrix behavior remains linear elastic and that all the inelastic deformations are localized along existing discontinuities that can re-activate in shear and tensile failure in response to fluid injection.
We neglect inertia as the growth of fluid-driven ruptures is slow (sub-meter per second velocity), although frictional instabilities can occur \citep{GaGe12}. We resolve these instabilities via quasi-dynamic damping \citep{Rice93}. In the sequel, we restrict to the cases where the rock matrix can be assumed impermeable at the scale of the injection duration - a limit often encountered in practical applications in low permeability rocks. This choice allows aims to highlight more explicitly some of the important features of the fluid-driven rupture growth problem. 
We first discuss the necessary physical hydro-mechanisms that must be accounted for, especially with regards to the constitutive behavior of the discontinuities. We present the details of the solver which combines boundary element for quasi-static elasticity and finite element for non-linear pore-fluid diffusion. A elastic predictor - plastic corrector scheme solves consistently the constitutive interface relation using the displacement discontinuity and pore-pressure as the primary unknowns. As usual, efficient block pre-conditioning of the tangent hydro-mechanical system is critical to achieve, in combination with appropriate adaptive time-stepping, adequate computational performance as  discussed on some rupture propagation examples. 
More importantly, we demonstrate accuracy on a consistent series of verification tests of increasing complexity for  3D fluid-driven frictional ruptures as well as tensile mode hydraulic fractures propagation. Computational efficiency is further discussed on two realistic configurations involving several discontinuities.

\subsection*{Notation}
We use a Cartesian frame throughout defined by the basis vector $\v{e}_{i},\,i=1,\,2,\,3$,
and denote $\v{x}=x_{i}\v{e}_{i}$ ($\v{y}=y_{i}\v{e}_{i}$ $)$ as the coordinates vector. 
For convenience, we either use indices notation (with the usual summation convention over repeated indices), e.g. denoting $\sigma_{ij}$ for the stress tensor, or bold-face for tensor and vectors. Similarly, either subscripts comma, e.g. $f_{,i} = \partial f/\partial x_i$, or 
operator, $\pmb{\nabla} f =   \partial f/\partial x_i \pmb{e}_i $, notation are used for spatial derivatives. The use  being clear depending on the context.  We refer to time as $t$ and write the time-derivative of $f$ explicitly as $\dfrac{\partial f}{\partial t}$ or alternatively as $\dot{f}$ for short.
 
\section{Problem Formulation}

Accounting for an initial stress state $\bm{\sigma}^o$ in equilibrium with gravity and far-field tectonic loads, the quasi-static balance of momentum of the rock mass can be re-written as follows:
\begin{eqnarray}
\bm{\nabla} \cdot \left(\bm{\sigma}-\bm{\sigma}^o \right)
=0\qquad\text{in } \Omega \label{eq:mech_p_1} \\
T_i=\bm{\sigma} \cdot \bm{n} \cdot \bm{e}_i=T^{g}_i\qquad\text{on }\Gamma_{T_{i}}\\
u_{i} =u_{i}^{g}\qquad\text{on }\Gamma_{u_{i}}\\
\Gamma_{u_{i}}\cap\Gamma_{T_{i}}=\emptyset & \qquad\Gamma_{u_{i}}\cup \Gamma_{T_{i}}=\Gamma 
\label{eq:mech_p_2}
\end{eqnarray}  
where $\Gamma $ denotes the overall boundaries of the domain,  $\Gamma_{u_{i}} $  and $\Gamma_{T_{i}}$ are the non-intersecting parts of the solid boundary where the displacement, respectively the tractions are imposed. 
We will distinguish between the pre-existing fracture(s) boundaries $\Gamma_{f}$ in the inner part of the domain and the outer domain boundaries $\Gamma_{\partial \Omega}$ (which possibly extend to infinity), such that $\Gamma = \Gamma_{f} \cup \Gamma_{\partial \Omega}$.  
$T_{i}=\sigma_{ij}n_{j}$ denotes the  traction vector acting on a facet of normal $n_i$. 

We focus on problems driven by fluid injection at a specific location either under a given volumetric flow rate or controlled in terms of fluid over-pressure (above the hydrostatic). The associated pore-fluid pressurization will trigger ruptures on existing discontinuities (fractures, faults) - either via shear failure  or a combination of shear and tensile failure. Our interest lies in the modeling of the growth of these ruptures under arbitrary fluid injections. The irreversible deformation will be strictly localized along existing discontinuities.

\subsection{Behavior of the rock matrix}
We assume that all the non-linearities will occur on the pre-existing discontinuities while the rock matrix behaves linearly (with
its pore-space saturated with fluid).
The constitutive relation (including pre-stress and initial pore-pressure) for a linearly isotropic porous solid matrix can be written in terms of stress ($\sigma_{ij}$) - strain ($\epsilon_{ij}$), pore-pressure ($p$) and porosity ($\phi$) changes as \citep{Cous04}:
\begin{eqnarray}
	\sigma_{ij} - \sigma_{ij}^o = 2 G \epsilon_{ij} + \frac{2G \nu }{1-2\nu} \epsilon_{kk}\delta_{ij} - \alpha (p-p^o) \delta_{ij}  \\
	\varphi= \phi - \phi^o = \alpha \epsilon_{kk} + \frac{p-p^o}{N}
\end{eqnarray}
where $G$ and $\nu$ are the drained shear modulus and Poisson's ratio of the porous solid, $\alpha$ is the Biot's coefficient and $N$ the intrinsic Biot modulus. The superscript $^o$ denotes the value in the initial reference state.
The fluid mass per unit of volume of porous media $m_f=\rho_f \phi $ can change due to porosity and fluid density variations. 
Defining the variation of fluid content as $\zeta = (m_f - m_f^o) / \rho_f^o  $, for a fluid compressibility $c_f$,  linear poroelasticity entails the following constitutive relation: 
\begin{equation}
	\zeta = (m_f - m_f^o) / \rho_f^o = (\phi-\phi^o) + c_f \phi^o (p-p^o) = \alpha \epsilon_{kk} + \frac{p-p^o}{M}
	\label{eq:zeta_bulk}
\end{equation}
where the inverse of the Biot's modulus $1/M$ is the sum of the pore mechanical ($1/N$) and fluid ($\phi^o c_f$) compressibilities.
The conservation of pore-fluid mass per unit of volume of porous media $m_f$ reduces to first order for a slightly compressible liquid to the following volume continuity equation (see e.g.\cite{RiCl76,DeCh93,Cous04}):
\begin{equation}
	\frac{1}{\rho_f^o}\frac{\partial m_f}{\partial t} + \bm{\nabla} \cdot \bm{q}  = \gamma 
	\label{eq:Fluid_volume_conservation}
\end{equation}
where  $\gamma $ is a source term and $q_i$ is the fluid discharge vector given by Darcy's law (Darcy's velocity):
\begin{equation}
	\bm{q} = -\frac{k}{\mu_f} \left( \bm{\nabla}p - \rho_f \bm{g} \right)
\end{equation}
where $k$, $\mu_f$, and $\rho_f$ denote the intrinsic permeability of the porous solid, fluid viscosity, and density respectively. $g_i$ is the Earth's gravity vector. 

 \begin{figure}
 \begin{center}
 \includegraphics[width=0.7\textwidth]{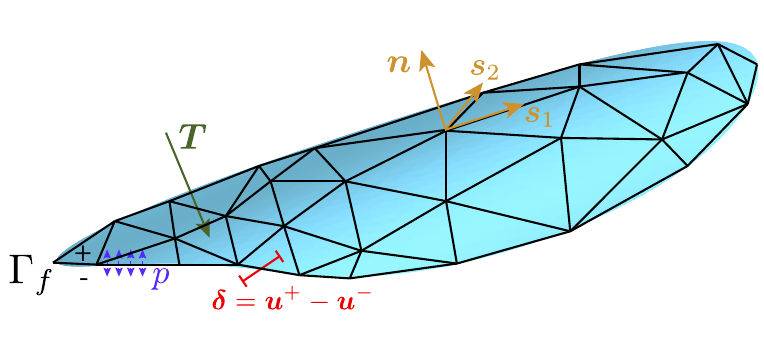}
 	\caption{Schematic of a discontinuity $\Gamma_f$ filled with fluid at pressure $p$, with the 
    definition of the local cartesian frame 
    $(\bm{s}_{1},\bm{s}_{2},\bm{n})$, traction  and displacement discontinuity vectors. Partial triangulated mesh as example.
 	\label{fig:Notation-interface}}
 	\end{center}
 \end{figure}
 
We {\em solely} focus our discussion on the limiting case where the rock matrix can be considered impermeable at the scale of the duration of the fluid injection. This limit is commonly encountered in practical engineering applications, as fractures typically have a permeability much larger than the rock matrix. In this limit,  matrix poroelasticity reduces to an elastic problem with undrained properties, i.e. the bulk elastic moduli $K$ is simply replaced by $K+\alpha^2 M$ (as the variation of fluid content is null in the rock matrix, such that \eqref{eq:zeta_bulk} reduces to $(p-p_o)=- M \alpha \epsilon_{kk} $. The restriction to the case of an impermeable matrix allows to focus on the important aspects associated with 
 non-linearities of fluid flow in the discontinuities and the associated numerical difficulties. It also stems from the fact that the 
  semi-analytical solutions for fluid-driven ruptures that have been derived in recent years 
    are restricted to such a case of an impermeable rock matrix. 
For tensile rupture (hydraulic fractures), some existing solutions account in a simplified manner fracture-matrix fluid exchange via early-time approximation (via one-dimensional perpendicular leak-off models for example \citep{HoFa57,KaDo20}). 
In the remaining of this paper, the matrix is assumed impermeable/elastic with elastic properties $G$ and $\nu$. Our aim being to clearly focus on the most non-linear aspects of the problem.
We briefly discuss the extensions to include matrix diffusion and poroelasticity in conclusions. 

\subsection{Hydro-mechanical behavior of discontinuities: fractures \& faults \label{sec:interface}}
	
Recognizing that rock joints / fractures as well as faults can be considered at the macroscopic scale of interest as pre-existing discontinuities. We model them at macroscopic scale as zero thickness interfaces but recognize at the microscopic scale their non-zero thickness $w$ notably to in relation to fluid flow.  We use interchangeably the term interfaces, fracture, faults or discontinuity throughout.  

 For a discontinuity with a local normal $\bm{n}$, we define locally at
any point along the discontinuity, an orthonormal Cartesian frame defined
by the unit vectors  
$(\bm{s}_{1},\bm{s}_{2},\bm{n})$
where $\bm{s}_{1}$ and $\bm{s}_{2}$ define a 2D frame in-plane of the interface locally tangent to
the interface (see Fig.~\ref{fig:Notation-interface}). 
The traction and displacement discontinuity vectors in this local frame are written as: 
\begin{align*}
\bm{T} & =T_{s_{1}}\bm{s}_{1}+T_{s_{2}}\bm{s}_{2}+T_{n}\bm{n}\\
\bm{\dd} & =\bm{u}^{+}-\bm{u}^{-}=\dd_{s_{1}}\bm{s}_{1}+\dd_{s_{2}}\bm{s}_{2}+\dd_{n}\bm{n}
\end{align*}
where the $n$ and $s$'s subscripts therefore corresponds to the normal and shear components. 
In view of their slenderness, fractures and faults are self-equilibrated, such that the action-reaction law apply (the traction vector is continuous across the discontinuity):
\begin{equation}
	T_i^+ + T_i^- = (\sigma_{ij}^+ - \sigma_{ij}^-) n_j^+ =0 
\end{equation}   
  where the positive and negative superscript denote the relative sides of the interface defined with outward normals (see Fig.\ref{fig:Notation-interface}). 
	
\subsubsection{Fluid flow}
\begin{figure}
 \begin{center}
 \includegraphics[width=\textwidth]{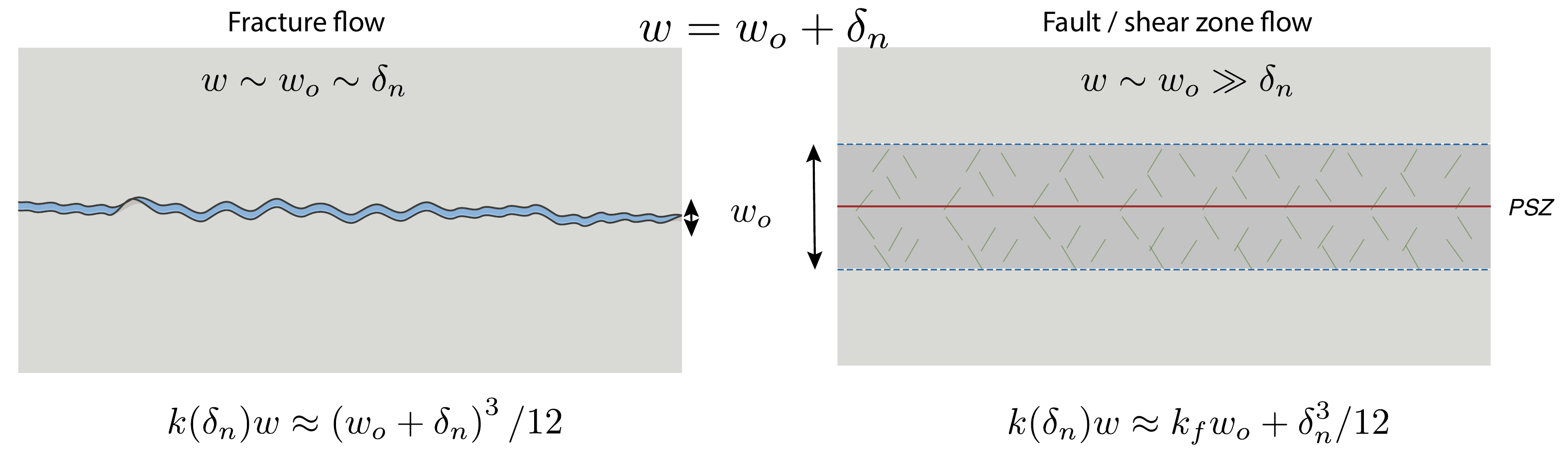}
 	\caption{Differences between a fracture (left) and a fault/shear zone (right) with a principal slip zone (PSZ). Fluid flow can occurs over a larger thickness than any induced dilation of the principal slip zone for a fault. On the contrary, for a fracture, flow is mostly restricted to the locus of the displacement discontinuity. It results in different storage and evolution of hydraulic transmissibility $k w$ as function of dilatant opening.
 	 	\label{fig:fracture-vs-faultzone}}
 	\end{center}
 \end{figure}

The main difference between a bare fracture and a fault resides in the presence of gouge materials and a damage zone in the latter (see Fig.~\ref{fig:fracture-vs-faultzone}). 
We model a fault as an interface where slip is localized in a principal slip zone, but where fluid flow takes place over a zone of larger thickness (including its damage zone). Bare fractures on the other hand, although having asperities at micro-scale, hold fluid flow strictly within the locus of the displacement discontinuity. Of course, a wide range of configurations exist between these two limits. A similar modeling framework holds for both cases pending the use of adequate models for the evolution of permeability, and proper constitutive parameters.   

Denoting $w$ as the thickness of the interface over which fluid flow occurs, the width integrated fluid mass balance for a slightly compressible liquid \eqref{eq:Fluid_volume_conservation} reads:
\begin{eqnarray}
\frac{1}{\rho_f^o}\frac{\partial w m_f}{\partial t} + \nabla_{||} \cdot(w \v{q}_{||}) + q_{\perp}^+ + q_{\perp}^-  = w \gamma  \label{eq:interface_conservation} \\
\v{q}_{||} = - \frac{k}{\mu_f} \left( \v{\nabla}_{||} p  - \rho_f \v{g}_{||} \right)   \label{eq:interface_Darcy}
\end{eqnarray} 
where the subscript $||$ restricts the component of vectors and differential operators to the coordinates in the fracture mid-plane (locally in the tangent directions $\bm{s}_1,\bm{s}_2$). More precisely, the along the fracture fluid discharge $q_{||}$ corresponds to the width-averaged Darcy's fluid velocity. 
For quiescent initial conditions ($\v{q}_{||}(t=0)=\bm{0} $), the initial pore-pressure $p^o$ is hydrostatic such that $\v{\nabla}_{||} p^o  = \rho_f \v{g}_{||} $, such that Darcy's law can be re-expressed as function of the over-pressure above hydro-static $ p-p^o$: $\v{q}_{||} = - \frac{k}{\mu_f} \v{\nabla}_{||} (p-p^o) $.

The leak-off velocities from the upper $ q_{\perp}^+$  and $ q_{\perp}^-$ account for fluid exchange with the rock matrix. It can be accounted for by directly coupling the interface and matrix flow explicitly \citep{BeDo19b} or in the view of the short time-span of fluid injection via approximate 1D diffusion models in the matrix in a direction perpendicular to the discontinuity \citep{KaGa19,KaDo20}. In what follows, for clarity, we focus on the impermeable case, and drop the fluid exchange between the rock and the matrix. 

\subsubsection{A poroelastoplastic interface law}

We model faults and fracture as zero thickness poroelastoplastic interfaces for which at microscale, a filling porous material (gouge) and/or asperities are present.
For existing discontinuities such as rock joints, a large number of models 
\citep{GeCa90,CaPr97} introduces a split between an elastic part of the displacement discontinuity (controlled by interface stifnesses) and a 'sliding'/ 'opening' part activated when the traction reaches a frictional yield limit. Such a sliding and opening parts of the displacement discontinuity is sometimes referred to as plastic, following the classical terminology of elasto-plasticity, although it can be reversible via reversal of the loads. It is more accurate to refer to inelastic or crack like displacement discontinuity for such an interface. 
An elasto-plastic like formalism (also accounting for damage) allows to properly modeled the deformation of rock joints, accounting for the important effect of variable dilatancy via the use of a non-associated flow rule. Such formalism is very similar to cohesive zone models (CZM) of fracture \citep{PaPa13}, although the interface stiffness is in reality for rock not a mere penalty parameter but allows to properly model the non-linearity of contact closure \citep{BaLu83}.

We decompose the displacement discontinuity vector across the interface in an elastic $\dd_{i}^{e}$ (reversible)
and inelastic / crack-like $\dd_{i}^{p}$ part:
\begin{equation}
\dd_{i}=\dd_{i}^{e}+\dd_{i}^{p}.
\label{eq:dd_split}	
\end{equation}
The variation of porosity of the interface $\phi-\phi^o=\varphi$ is split in a similar fashion ($\varphi = \varphi^e+\varphi^p$).  
The increment of plastic/inelastic work of the interface (of thichkness $w$) is thus: $\text{d} W^p = T_i \text{d}\dd_{i}^{p} + p \text{d}(w \varphi^p) $. Like for all geomaterials, the variation of plastic porosity is equal to the plastic volumetric strain as the solid constituents are plastically incompressible in comparison to the grain re-arrangements leading to porosity changes \citep{Cous04}. For an interface, this translates to  $  \dd_n^p = w \varphi^p $. We therefore obtain that the increment of plastic work is a function of the Terzaghi's effective tractions $\bm{T}^\prime=\bm{T}+p\bm{n}$:
\begin{equation}
	\text{d} W^p = ( T_i + p \,n_i) \text{d} \dd_i^p = T^\prime_i \text{d} \dd_i^p .
\end{equation}
All experimental observations in geomaterial (matrix and discontinuities) confirm that Terzaghi effective stress indeed drives inelastic deformation both for matrix and interface inelastic deformation \citep{Cous04}. 

Accounting for inelastic displacement discontinuity, upon  integration over the width $w$, the poroelastoplastic constitutive relation of the interface can be expressed in terms of the traction $\bm{T}$ and the elastic part of the displacement discontinuity vector  $\bm{\dd}^e = \bm{\dd} - \bm{\dd}^p$:
\begin{eqnarray}
	\bm{T}-\bm{T}^o = \bm{\mathbb{C}} \cdot \left( \bm{\dd} - \bm{\dd}^p \right) - \alpha (p-p^o) \bm{n} \\
	\varphi - \varphi^p  =  \alpha \frac{(\dd_n - \dd_n^p )}{w} + \frac{p-p^o}{N}
\label{eq:interface_poroel}
\end{eqnarray}
In the previous constitutive relation, $ \bm{\mathbb{C}} $ denotes the interface stiffness matrix (in $N/m$) which is possibly non-linear (as function of the Terzaghi's effective tractions). In the local cartesian reference frame of the interface, assuming an isotropic behavior, $ \bm{\mathbb{C}} $  is written as function of a normal $K_n$ and shear $K_s$ stiffness:
\begin{equation}
 \bm{\mathbb{C}}=K_{s}\left(\bm{s}_{1}\otimes\bm{s}_{1}+\bm{s}_{2}\otimes\bm{s}_{2}\right)+K_{n}\bm{n}\otimes\bm{n}
\end{equation}

For usual 'filling' materials inside the discontinuity (granular gouge or/and asperities), the porous skeleton is much more compliant than its solid constituents such that the Biot's coefficient of the interface reduce to unity: $\alpha=1$ \citep{DeCh93,Cous04}. As a result, like for soils, the Biot's and  Terzaghi's effective tractions coincides (the former strictly governing the poroelastic deformation while the later drives plastic deformation - see \cite{Cous04} for discussion). In this limit, the intrinsic Biot pore compressibility $1/N$ is usually negligible compared to the fluid compressibility. We assume $\alpha=1$, and keep the possibility of non-zero pore compressibility of the filling material. 
It is also important to note that if the interface/fracture fails in tension, upon full opening, the shear and normal stiffnesses of the interface vanishes (and the effective traction vector is stritcly zero) as discussed in the next subsections.  

	After re-arranging \eqref{eq:dd_split}-\eqref{eq:interface_poroel} with the previous assumptions, the  constitutive relations of the poroelastoplastic interface reduces to:  
\begin{eqnarray}
	\bm{T}^\prime-\bm{T}^{\prime\,o} = \bm{\mathbb{C}} \cdot \left( \bm{\dd} - \bm{\dd}^p \right)  \\
	 w\varphi = \dd_{n}+ \frac{w}{N} (p-p^o) 
	 \label{eq:interface-porosity-variation}
\end{eqnarray}
The width integrated variation of fluid content of the interface is $ w (m_f-m_f^o)/\rho_f^o = w\varphi  + w  \phi^o c_f (p-p^o)$. Note that the interface width $w$ is the sum of an initial thickness $w^o$ and the normal displacement discontinuity: $w=w^o+\dd_n$.  The width integrated fluid conservation \eqref{eq:interface_conservation} on the interface can thus be rewritten as
\begin{equation}
	\frac{\partial \dd_n }{\partial t} +\frac{w}{M} \frac{\partial p}{\partial t} + \nabla_{||} \cdot(w \v{q}_{||}) + q_{\perp}^+ + q_{\perp}^-  = w \gamma 
	\label{eq:interface_conservation_final}
\end{equation}
with $1/M=1/N+\phi^o c_f$ the Biot's modulus of the fluid saturated porous interface.

\subsubsection{Interface yield criteria, flow rule, and damage \label{subsec:Interface_plasticity}}

Inelastic displacement discontinuity occur when a yield criteria written
in term of the effective tractions acting on the interface is reached.
Besides effective tractions, the yield criteria may also depend on
one or more internal variables $\bm{\chi}$ tracking the internal
state of the interface (such as the cumulated plastic shear slip).
Denoting $F(\bm{T}^{\prime},\bm{\chi})$ the yield function, sole
elastic deformation occur when $F(\bm{T}^{\prime},\bm{\chi})<0$,
while the constraint $F(\bm{T}^{\prime},\bm{\chi})=0$ is active upon
plastic yielding. 

\begin{figure}
    \centering
    \includegraphics[width=0.8\linewidth]{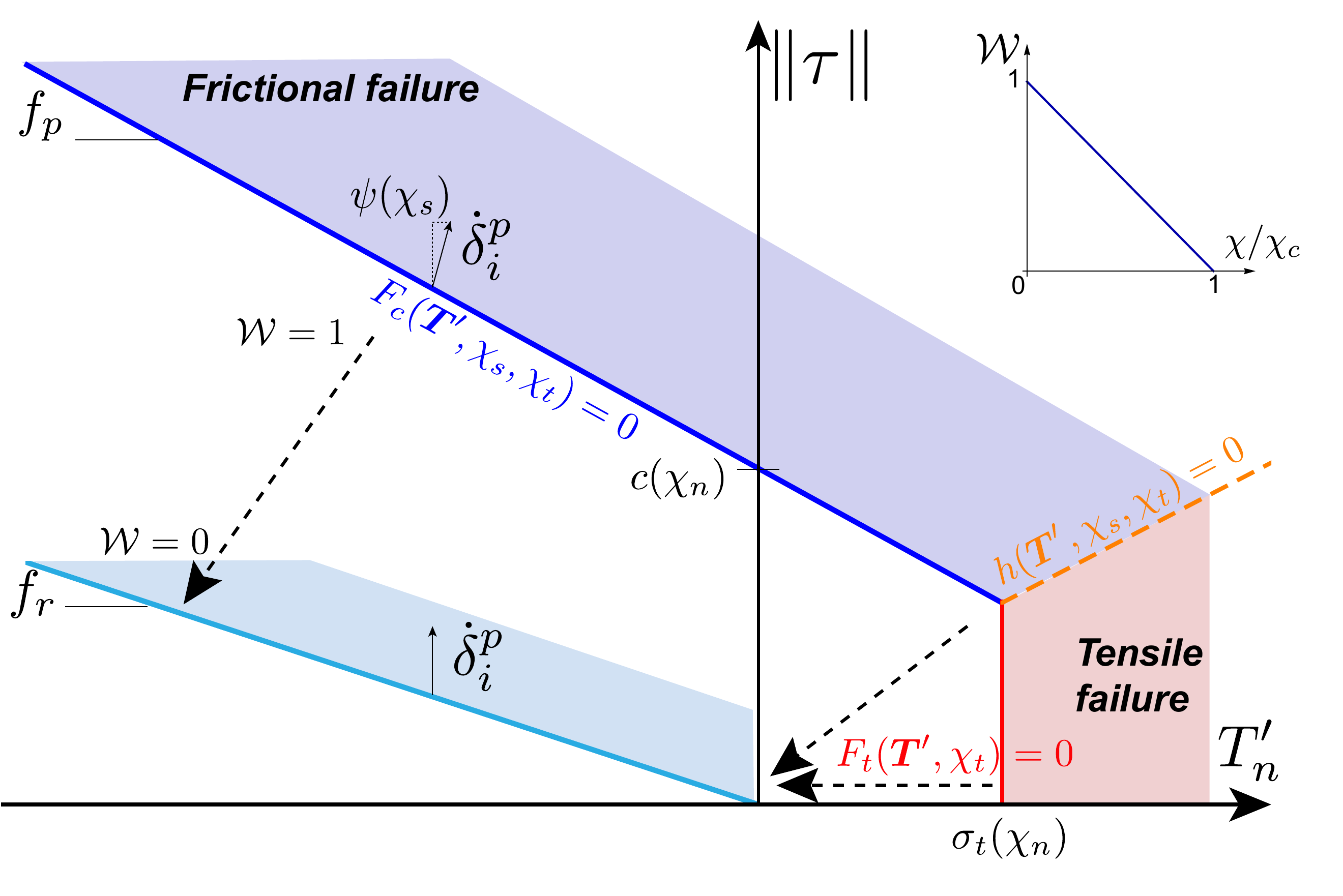}
    \caption{Combined yield functions for Coulomb frictional and tensile failure with softening for a pre-existing interface as function of effective tractions. Softening evolves with accumulated slip (friction) and total plastic displacement discontinuity (cohesion and tensile strength) to reach an ultimately cohesionless residual Coulomb failure with no more dilation.} 
    \label{fig:yield}
\end{figure}

To capture with the simplest possible model, most of the mechanisms observed experimentally for rock joints and fractures \citep{Bart76,BaBa85}, we combine a Coulomb yield surface with a tensile cut-off and allow for the weakening of cohesion, friction and tensile strength (see Figure \ref{fig:yield}). The combination of these two distinct criteria notably allows to simulate the observed difference in both shear and tensile fracture of rock joints. 
A non-associated plastic flow rule is used for the plastic displacement related to Coulomb failure in order to properly model shear-induced dilatancy and the transition to a critical state, while an associated flow rule is used for the tensile cut-off.
Other choice of constitutive models is of course possible - see \cite{GeCa90,CaPr97,MrGi96,StMr01,SoLe04,ZaAl13,LiMi16,LiWu20} among many others.   
In the local frame of reference of the interface, it is useful to introduce the shear component of the effective traction $\bm{\tau}$ and the principal shear direction vector $\bm{\hat{s}}$:
\begin{equation}
\bm{\tau}=T^\prime_{s_{1}}\bm{s}_{1}+T^\prime_{s_{2}}\bm{s}_{2}=\|\bm{\tau}\|\bm{\hat{s}}\qquad\bm{\hat{s}}=\frac{\bm{\bm{\tau}}}{\|\bm{\bm{\tau}}\|}\qquad\|\hat{\bm{s}}\|=1\qquad\bm{\tau}\cdot\hat{\bm{s}}=\|\bm{\bm{\tau}}\|	
\end{equation}
 $\hat{\bm{s}}$ is an unit vector
indicating the principal direction of shear, with of course 
$\bm{n}\cdot\hat{\bm{s}}=0$ and the effective traction can be written as
\[
\bm{T^{\prime}}=\bm{\tau}+T_{n}^{\prime}\bm{n}.
\]
Similarly, we define the shear slip vector as 
\[
\bm{\delta}_s=\dd_{s_{1}}\bm{s}_{1}+\dd_{s_{2}}\bm{s}_{2} =\dd_s \hat{\bm{s}}
\]
such that the displacement discontinuity vector is $\bm{\dd}=\dd_s \hat{\bm{s}}+\dd_{n}\bm{n}$.
We track the state of the interface with two distinct internal variables $ \bm{\chi}=(\chi_t,\,\chi_s)$: 
\begin{enumerate}
    \item The maximum accumulated total inelastic displacement discontinuity magnitude 
    \[ \chi_{t}=\max(\|\bm{\dd}^{p}(t)\|) \] to model the evolution of cohesion and tensile strength,
    \item the maximum accumulated inelastic shear slip displacement discontinuity magnitude 
    \[\chi_{s}=\max(\|\bm{\delta}^{p}(t)\|) \] to model the evolution of the friction and dilation coefficient of the discontinuity.
\end{enumerate}

\paragraph{Coulomb failure \\}
The Coulomb yield function for an interface is given by
\[
F_{c}(\bm{T}^\prime,\bm{\chi})=\|\bm{\tau}\|+f(\chi_s)T^\prime_{n}-C(\chi_n)=\bm{T}^\prime\cdot\hat{\bm{s}}+f(\chi_s)\bm{T}^\prime\cdot\bm{n}-C(\chi_n)
\]
with possibly evolving friction coefficient $f$ and cohesion $C$. 
Upon reaching this frictional yield limit, inelastic variation of displacement discontinuity occurs according to a non-associated flow rule:
\begin{align*}
F_{c}(\bm{T}^{\prime},\bm{\chi})<0 & \qquad\dot{\bm{\dd}}^{p}=0\\
F_{c}(\bm{T}^{\prime},\bm{\chi})=0 & \qquad\dot{\bm{\dd}}^{p}=\dot{\lambda}\bm{\nabla}_{\bm{T}^{\prime}}G\qquad\dot{\lambda}>0
\end{align*}
where a dot $\dot{\square}$ denotes the time-derivative. 
$G(\bm{T}^{\prime},\bm{\chi})$ is a non-associated inelastic flow potential taken here as 
\[
G(\bm{T},\bm{\chi})=\|\bm{\tau}\|+\psi(\chi_s)T^\prime_{n}
\]
where $\psi\ge0$ is a dilatancy coefficient which evolves as function of the internal variable controlling the evolution of friction. The dilatancy coefficient decreases as inelastic slip accumulates such that the interface eventually reaches a criticial state where no inelastic dilation occur, i.e. $\lim_{\chi_s\rightarrow \infty } \psi =0$.
 
The inequality related to the activation of the yield criteria and the inelastic deformation has the following complementary Karush-Kuhn-Tucker
condition 
\[
\dot{\lambda}F_{c}=0
\]
which reflect the fact that if $F_{c}<0$, $\dot{\lambda}=0$ no inelastic displacement discontinuity 
occur while if $\dot{\lambda}>0$ then the yield function must be
satisfied $F=0$. 
The gradient of $F_{c}$ and $G$ with respect to the effective traction $\bm{T}^\prime$ are chiefly obtained as:
\begin{align*}
\bm{\nabla}_{\bm{T}^\prime}F_{c} & =\frac{\bm{\bm{\tau}}}{\|\bm{\tau}\|}+f(\chi_s)\bm{n}=\hat{\bm{s}}+f(\bm{\chi})\bm{n}\\
\bm{\nabla}_{\bm{T}^\prime}G & =\hat{\bm{s}}+\psi(\chi_s)\bm{n}=\bm{N}(\bm{T}^\prime)
\end{align*}
such that the non-associated Coulomb inelastic flow rule can be re-written as:
\[
 F_{c}(\bm{T}^\prime,\bm{\chi}) =0 \qquad \dot{\bm{\dd}}^{p}=\dot{\lambda}(\hat{\bm{s}}+\psi(\chi_s)\bm{n})  \qquad  \dot{\lambda}>0.
\]
The plastic multiplier $\dot{\lambda}$ is function of the global equilibrium and locally obtained from the
condition that during an inelastic evolution, the state of effective tractions must remain
 on the yield surface, in other words when $\dot{\lambda}>0$, we must
 have 
 \[
 \dot{F}_{mc}=\bm{\nabla}_{\bm{T}^{\prime}}F_{c}\cdot\bm{\dot{T^{\prime}}}+\bm{\nabla}_{\bm{\chi}}F_{c}\cdot\bm{\dot{\chi}}=0
 \]
 which is referred to as the consistency condition in the theory of elasto-plasticity.

\paragraph{Evolution of friction, dilatancy and cohesion \\}
 
The friction and dilation coefficients are taken function of the maximum accumulated
inelastic shear displacement discontinuity $\chi_s$ \citep{Ida72,PaRi73}. 
The friction and dilation coefficients are assumed to vary from an
initial value $f_{p}$ (respectively $\psi_{p}$) to a residual value
$f_{r}$ (respectively to $\psi_{r}=0$ - reproducing the fact that frictional sliding
ultimately reaches a critical state without any more dilation). 
We write the following evolution 
\begin{align*}
f(\chi_s) & =\left(f_{p}-f_{r}\right)\mathcal{W}(\chi_s/d_{c})+f_{r}\\
\psi(\chi_s) & =\psi_{p}\mathcal{W}(\chi_s/d_{c})
\end{align*}
with $\mathcal{W}(x)$ a smooth function going from 1 (at $x=0$) to zero (at $x \ge  1$).
 $d_{c}$ is the critical
slip weakening distance scale. 
The simplest choice for $\mathcal{W}$ is a linearly decreasing function between $0$ and $1$.
	
The cohesion is taken function of the maximum total accumulated inelastic
 displacement discontinuity $\chi_t$. It typically decreases from a peak value $c_p$ to zero over
 a critical distance of inelastic displacement discontinuity $w_c$: 
\[
c(\chi_{t})=c_p \mathcal{W}(\chi_{t}/w_{c}).
\]
The critical opening distance $w_c$ may be different than the critical slip distance $d_c$, reflecting a different critical fracture energy in tension and shear failures. 

\paragraph{Tensile failure \& contact \\}
In addition to Coulomb frictional failure with cohesion, we add a tensile cut-off 
to better capture the failure of fracture in tension (notably to simulate the creation of new fractures).  The yield criteria in tension is simply expressed as: 
\begin{equation}
F_{t}(\bm{T}^\prime,\bm{\chi})=T^\prime_{n}-\sigma_{t}(\chi_{n})=\bm{T}^\prime\cdot\bm{n}-\sigma_{t}(\chi_{n})\le 0.
\label{eq:tensile_yield}	
\end{equation}
The flow rule for such a pure tensile cut-off is 
\begin{equation}
F_{t}(\bm{T}^\prime,\bm{\chi})=0 \qquad \dot{\bm{\dd}}^{p}>0
\label{eq:tensile_flow}
\end{equation}
with the complementary condition $F_{t}(\bm{T}^\prime,\bm{\chi}) |\dot{\bm{\dd}}^{p} |=0 $.
The  inelastic displacement discontinuity depends on the loading/unloading sequence from the application of the constraint $F_{t}(\bm{T}^\prime,\bm{\chi})=0$ in the solution of the global equilibrium.

The evolution/degradation of the tensile strength as function of the
maximum encountered 'inelastic' displacement discontinuity $\chi_t$ is written similarly than for the evolution of cohesion:  
\[
\sigma_{t}(\chi_{t}=\max(\|\bm{\dd}^{p}\|))= \sigma_p \mathcal{W}(\chi_t/w_c)
\]
with $\mathcal{W}$ a linearly decreasing function from 1 (when $\chi_t/w_c=0$) to 0 at $\chi_t/w_c=1$. 
If the yield criteria is satisfied $F_t(\bm{T},\bm{\chi})=0$, neglecting
the elasticity of the interface (rigid prior to yield case), we obtain
a classical mode I traction-separation law between 
normal traction and opening of cohesive zone models. 
 
We define a damage measure of the interface as $d=\min(\chi_t/w_c,1) $. Upon full damage, when the tensile strength of the interface is now null,  the elastic stiffness of the interface vanishes if the tensile failure criterion is active. In other words,
\begin{equation}
 K_n=K_s = 0 \text{ if } F_{t}(\bm{T}^\prime,\chi_n\ge w_c)=0 	
\end{equation}

If the interface opens as a crack (fully damaged), upon unloading, contact will occur. We model the fact that at micro-scale the created fracture exhibit a roughness by introducing a positive macroscopic inelastic residual width $w_R$. 
The Signorini contact conditions when the interface has a completely degraded tensile strength therefore reads:
\begin{eqnarray}
 \dd_n^p  - w_R \ge 0  \qquad  F_{t}(\bm{T}^\prime,\chi_n\ge w_c) \le  0  \qquad  ( \dd_n^p  - w_R ) F_{t}(\bm{T}^\prime,\chi_n) =0
\end{eqnarray} 
Under pure mode I loading, the residual width $w_R$ upon full damage, is taken as a fraction of the critical width $w_c$: $w_R = \alpha w_c$ (with $\alpha \le  1$ ).

We have written the contact condition as function of the inelastic part of the displacement discontinuity in order to ensure a positive total displacement discontinuity (no overlap) under compression. This is particularly important as the non-linearity of the interface flow properties are related to the normal component of the displacement discontinuity.  
It is important to note that
for a pre-existing fractures $\sigma_p=0$, we must have $w_R=0$ ($w_c=0$)   in this formulation. 
This can be grasped by the following example. Assuming an initial normal compressive traction as $T^\prime_n = -\sigma_o $, under pure tensile loading, the interface will open when the elastic normal opening (from the initial state) is $\dd_n^{e,o}=  \sigma_o /K_n $.  For such a pre-existing fracture that opens, upon unloading, contact starts when the total normal displacement discontinuity goes back to  $\dd_n^{e,o}$ or alternatively when the inelastic part of the normal  displacement discontinuity goes to zero, therefore indicating that $w_R=0$ under pure mode I loading for pre-existing / fully broken / fractures.

 Of course, the  residual width $w_R$ also evolves with inelatic shear-induced dilation under frictional failure due to shear loading. The evolution of the residual width $w_R$ must therefore be written as: 
\begin{equation}
	w_R = \alpha w_c + \int_0^t \psi(\chi_s) \chi_s \text{ d}t
\end{equation}  
to account for the effect of shear-induced dilation in mixed mode loading.
This notably allows to model the case where shear-induced dilation is followed by tensile opening and then closure.

\paragraph{Effective stress dependent stiffness \\}

The interface stiffness are only active when the interface is not yielded in tension. Notably for pre-existing / fully broken fractures, these elastic stiffness are only active when the interface is under compression effective stresses. 
Although from a modeling perspective, a constant interface stiffness is the simplest choice, rock interface/joints exhibit a clear non-linear response upon closure \citep{BaLu83,BaBa85}. 
We settle either for a constant value of normal and shear stiffnesses or for the well-accepted hyperbolic model first proposed by \citet{BaLu83} for the normal stiffness $K_n$. Such a phenomenological model - found adequate for un-mated rock joints - can be written as 
\begin{equation}
K_n  = K_i \left( 1 - \frac{T^\prime_n}{k_i v_m} \right)^2
\end{equation}
where $K_i$ is the stiffness upon closure (at zero effective normal traction) and $v_m$ is a characteristic closure distance. 
Note that $K_{i}v_{m}=\Sigma_{m}$
can be seen as a ``characteristic closure'' pressure: the stiffness
is four times the initial value when $-T_{n}^{\prime}=\Sigma_{m}$. For an interface with initially some tensile strength, one can account for the degradation of the normal stiffness due to tensile failure by decreasing the stiffness upon closure $K_i$ as function of the interface damage $d=\min(\chi_t/w_c,1) $, taking for example  $K_i =K_* \exp(-\beta  d)$  or other adequate functional forms. 
 
A linear dependence of the tangent stiffness on the normal effective stress has also been found to be adequate in some cases \citep{BaLu83,RuSt03}.  For uncorrelated fracture, the closure relation is found to be logarithmic
- see \citet{BaLu83}, and actually similar to well-known behavior
of the contact mechanics of rough uncorrelated surfaces \citep{Perr07}, and to the well-observed compaction relation of granular medium \cite{MiSo05}.  
In that case, the stiffness evolves linearly with $-T_{n}^{\prime}$ \citep{Corn15,BaLu83}: 
\[
K_{n}=-\kappa T_{n}^{\prime}
\]
such that the elastic variation of normal displacement discontinuity under compressive normal stress is logarithmic
\[
\dot{T}_{n}^{\prime}=\kappa T_{n}^{\prime}\dot{\dd_{n}}^{e}\rightarrow \dd_{n}=(1/\kappa)\log|T_{n}^{\prime}|/|T_{n}^{\prime o}|
\]

In comparison to the normal stiffness,
the non-linearities of the shear stiffness $K_s$ are not as well characterized. We settle for a constant value of $K_s$ for simplicity. This has notably the advantage of  simplifying the solution of the interface constitutive relation via the classical elastic-predictor plastic corrector scheme of elastoplasticity (see supplemental materials for details).  
It is also important to recall that upon mechanical opening of the pre-existing fracture, the loss of mechanical contact imposes that both the shear and normal stiffness goes to zero. A constraint that must be explicitly taken into account.

\paragraph{Selection of the failure mode and solution \\}

When the yield criteria for both failure modes are reached, we select 
an unique active  mode of failure based on a continuous function first proposed in \citep{FLAC} as:
\[
h(\v{T}^\prime,\v{\chi}) = \|\v{\tau}\|- (C(\chi_n)-f(\chi_s)\sigma_t(\chi_n)
+\alpha(\v{\chi}) (T_n^\prime - \sigma_t(\chi_n))
\]
with $\alpha(\v{\chi})=\sqrt{1+f(\chi_s)^2}-f(\chi_s)$. If $h>0$, shear failure is selected, otherwise tensile failure. The function $h$ separates the domain where shear and tensile failure is active - see Fig.~\ref{fig:yield}.

The rate of change of effective traction as function of the rate of change of the total displacement discontinuity of the interface is the solution of the previously described path-dependent non-linear elasto-plastic relations. 
It is solved using an elastic predictor - plastic corrector algorithm classical in elastoplasticity \citep{SiHu98,deSo11}.
The details of the integration of the interface constitutive description chosen here are given in Supplemental Materials. 
Such a local interface relation along the pre-existing discontinuities, must of course be solved in conjunction with the global quasi-static elastic equilibrium of the whole fractured medium and the fluid-flow along the discontinuities as discussed in section \ref{sec:interface}.

\subsubsection{Permeability evolution of the discontinuity}

In the formalism of coupled hydro-mechanics, for the case of closed
fractures/joints (under compressive effective stress), the evolution of permeability are sometimes expressed in terms of the current effective normal traction in a similar way than for granular soils \citep{Rice92}.
This is however no longer the case when the fractures mechanically open
- for which the effective normal traction becomes zero. It is more
convenient to express the variation of permeability as function of
the variation of the interface aperture, or the variation of inelastic interface porosity (the two being equal in our formalism). 
More specifically, it is the change in interface transmissibility $k w$ that governs the resistance to flow in the interface as described by eqs.(\ref{eq:interface_conservation})-(\ref{eq:interface_Darcy}).

\paragraph{Flow in bare fracture: modified cubic law \\
}
For mechanically open fractures, the normal displacement discontinuity is larger than the initial width of the interface $w_o$: $w=w_o + \dd_n \approx \dd_n$ as $\dd_n \gg w_o$. In that case, 
Poiseuille law directly provides the corresponding hydraulic transmissibility as $k w= \dd_n^3/12$. 

The formalism of the cubic law is also used for mechanically closed fractures (under compressive effective normal traction) \citep{Whit80}. However, it is  generalized by introducing a finite hydraulic width $\omega_{o}=f(w_{o})$ in addition to the mechanical aperture. This allows to capture the deviation from the cubic law at large compressive loads for which the mechanical aperture reduces to zero. 
The cubic law is thus written as function of the ``hydraulic aperture'' $\omega=\omega_{o}+\dd_{n}$ to
account for the fact that if initially closed, under significant compressive
stress, the interface  have a remaining  permeability. We write 
\begin{equation}
w k(\dd_n)=\frac{w (\omega_o + \dd_n)^2}{12}\qquad w=w_{o}+\dd_{n}
	\end{equation}
where $\dd_{n}$ is the normal component of the displacement discontinuity. In the reference configuration / initial state, it is necessarily equal to zero. This corresponds to the initial state for which the hydraulic transmissibility of the interface is $w_o \omega_{o}^2/12$.
The initial hydraulic width $\omega_o$ may possibly be different than the initial interface thickness $w_o$ \citep{ZiPa24}, although for bare fracture we settle for $\omega_o=w_o$.
A model that reproduces  well experimental data for the flow in fractures under compression effective stresses is given by:
\begin{equation}
	 w k(\dd_n) = \frac{1}{12} \left( \dd_n \times (1+\sigma_w/\dd_n)^{-3/2} +w_o \right)^3
\end{equation}
where $\sigma_w$  is akin to a roughness scale that governs the transition between a constant hydraulic transmissibility at small mechanical apertures to the classical cubic law at large mechanical apertures. All the simulations reported in this paper assumes $\sigma_w =0$ for simplicity.
In the absence of additional experimental data, we settle for the cubic law and otherwise needed takes for fractures the initial thickness and initial hydraulic aperture equal.  
\paragraph{Flow in fault/shear-zone \\
}
For the case of a thicker / mature fault zone, flow occurs within a unit containing a principal slip zone (see Fig.\ref{fig:fracture-vs-faultzone}).
 It differs from the case of a bare fracture, notably with respect to the amount of fluid stored - as here the total width of the flow $w$ remains greater than the opening component of the displacement discontinuity $\dd_n$. In a simplified model, the flow across the entire thickness of the fault zone  can be obtained by summing the flow within the damage/gouge zones and the principal localized slip zone (which corresponds to the locus of the displacement disontinuity). The overall hydraulic transmissibility of the "interface" can be written as:
\begin{equation}
	k w = k_f(T^\prime_n) w_o + k_{psz}(\dd_n) \dd_n 
\end{equation}
where $k_f$ is the permeability of the fault zone (which may be non-linearly dependent on the effective normal stress), and $k_{psz}$ is the permeability  of the principal slip zone associated with the displacement discontinuity.  For the latter, one can use the cubic law $k_{psz}(\dd_n) \dd_n = \dd_n^3/12$ for simplicity.  
\paragraph{Remark}
    The interface transmissibility (with respect to its initial value) for bare fractures and shear zones thus evolves slightly differently with respect to the initial state: in $(1+\dd_n/w_o)^3$ (for bare fractures) and $ 1+(\dd_n/(k_f w_o)^{1/3})^3$ (for shear zone) respectively. 
    More complex evolution of the fracture hydraulic transmissibility may be developed based on specific experimental results. All must capture the evolution from an initial value toward the cubic law when the interface mechanically opens. In the latter open state, the $\delta_n^3/12$ always dominates resulting in a very stiff coupling between fluid flow and mechanical deformation specific to hydraulic fracturing \citep{Deto16}.

\section{Numerical scheme}
	
We briefly describe a fully-coupled implicit scheme for the solution of the time-dependent problem of fluid flow along the fractures/faults combined with the balance of linear momentum of the medium accounting for the non-linear constitutive relations of the fractures/faults  previously described. The chosen methods aim at providing a robust, accurate and computational efficient solver for these  class of fluid-driven rupture problems. We combine well-known numerical methods for hydro-mechanical problems, elasto-plasticity and non-linear fluid flow.


\subsection{Solid mechanics}
	 Recognizing that all the non-linearites lies on the interface/fractures, we use a collocation boundary element method for the discretization of the balance of momentum for the linearly elastic rock domain. 
	 In particular, we use a displacement discontinuity method which is particular discretization of the traction hyper-singular boundary integral equation for linear isotropic quasi-static elasticity \citep{CrSt83,HiKe96,Bonn99,Mogi14}. 
	The traction hypersingular boundary integral equation can be written as:
	 \begin{eqnarray}
T_{i}(\bm{x}_{*}) -T_{i}^{o}(\bm{x}_{*})=\lim_{\bm{x}\rightarrow\bm{x}_{*}}\int_{\Gamma_{\partial \Omega}}\left[\mathbb{T}_{i}^{a}(\bm{y},\bm{x})\left(T_{a}(\bm{y})-T_{a}^{o}(\bm{y})\right)-\mathbb{H}_{i}^{a}(\mathbf{x},\mathbf{y})u_{a}(\bm{y})\right]\mbox{ d}S_{y}\nonumber \\ 
  -\lim_{\bm{x}\rightarrow\bm{x}_{*}}\int_{\Gamma_{f}}\mathbb{H}_{i}^{a}(\mathbf{x},\mathbf{y}) \dd_{a}(\bm{y})\mbox{ d}S_{y} 
\label{eq:TractionBIE-general}
\end{eqnarray}
where the fundamental elastic kernels appearing in this boundary integral
equation are:
\begin{itemize}
\item $\mathbb{T}_{i}^{a}(\bm{x},\bm{y})=S_{ij}^{a}(\bm{x},\bm{y})n_{j}$
the traction vector on a surface of normal $\bm{n}$ at $\bm{y}$
due to a unit point force in direction $\bm{e}_{a}$ located at $\bm{x}$
($\mathbb{S}_{ij}^{a}$ denotes the corresponding stress field),
\item $\mathbb{H}_{i}^{a}(\text{\ensuremath{\bm{y},\bm{x}})=\ensuremath{n_{j}}(\ensuremath{\mathbf{x}})\ensuremath{c_{ijkl}S_{ab,l}^{k}}(\ensuremath{\mathbf{y}},\ensuremath{\mathbf{x}})\ensuremath{n_{b}}(\ensuremath{\mathbf{y}})}$ 
corresponds to the traction vector on a surface at $\bm{x}$ due to a dislocation
dipole of normal $\bm{n}(\bm{y})$ and unit intensity located at $\bm{y}$.
It is hyper-singular.  
\end{itemize}
 
	 The boundary element method is particularly attractive to simulate injection at depth for which the domain of interest can be considered as infinite, which is our primary interest here. In that case, the first integral in \eqref{eq:TractionBIE-general} vanishes as the outer boundary  of the solid $\Gamma_{\partial \Omega}$ goes to infinity. 
	 The problem reduces to a single boundary integral equation between the traction and displacement discontinuity vector along the discontinuities $\Gamma_f$. 
	 We discretize the different fractures / faults and assume a piece-wise constant approximation of the displacement discontinuity over a given element, and collocate the boundary integral equation at the centroid of each element in the mesh \citep{CrSt83,HiKe96}. 	 
	 This choice simplify the treatment of the hyper-singular kernel. Notably, the integrals are evaluated before the limit is taken and as such remain regular. In particular, we use analytical integration of these integral as presented in \cite{HiKe96,Fata11,NiMo15} among many others. In what follow, we model either 2D plane-strain,  axisymmetric or fully 3D configurations - using respectively segment, ring and triangular surface elements for which these integrals are available analytically.

\citet{Rice93} introduced the concept of quasi-dynamic boundary-integral equations in the context of the modeling of earthquake cycles on faults. The approach retains only a local mass inertial
term (associated with $\rho\,\ddot{\bm{u}}$) to capture in a simple form wave radiation during a fast interface rupture. 
A local term $\bm{\eta}\cdot\dfrac{\partial\bm{\dd}(\bm{x},t)}{\partial t}$ is substracted to the quasi-static boundary integral operator, 
with 
\[
\bm{\eta}=\frac{1}{2}\left[\begin{array}{ccc}
G/c_{s} & 0 & 0 \\
0 & G/c_{s} & 0\\
0 &0  & G/c_{p}
\end{array}\right]
\]
where $c_{s}=\sqrt{\frac{G}{\rho}}$ and $c_{p}=\sqrt{\frac{K+4/3G}{\rho}}$ are the shear and compressional elastic
waves speed of the material. Such a quasi-dynamic term is active only at very large slip/opening rate, essentially only when an instability occurs in association to the weakening of the interface properties. It has merely no effects otherwise, but regularize fast interface ruptures. 

\paragraph{Final system}
Using a collocation displacement discontinuity
method, the  hypersingular boundary integral equations accounting for a quasi-dynamic term  reduces to a dense linear system 
\begin{equation}
	\bm{T}(t) - \bm{T}^o (t)= \bm{\mathbb{E}} \cdot \bm{\dd} (t)-\bm{\eta}\frac{\partial\bm{\dd}(t)}{\partial t}
	\label{eq:ElasSystem}
\end{equation}
where $\bm{T}$ is the vector of tractions at all collocation points 
in the mesh, $\bm{T}^{o}$ the vector of tractions at all collocation
points due to far-field in-situ stress (also possibly be
time-dependent), and $\bm{\dd}$ the vector of all displacement discontinuities at all collocation points at time $t$. $\bm{\eta}$ is a diagonal matrix containing the components of the
quasi-dynamics term for each collocation point, while $\mathbb{E}$ is the hypersingular elasto-static boundary element matrix. 
We leverage hierarchical matrix algorithms to significantly speed up the solution of the dense linear system associated with the boundary element method while lowering the memory requirements. The compression is achieved by applying low-rank approximations to far-field interactions, while near-field components are retained in their original form. We refer to \cite{BoGr03,Bede08,Hack15} for details on the theory and algorithms for hierarchical matrices. 
Our implementation follow closely the algorithm described in \cite{ChDe17}, 
and is available in open-source \citep{LeFa25}. 
The use of hierarchical matrix allows to lower the computation cost of the dot product   $\bm{\mathbb{E}}\cdot \bm{\dd} $ from $\mathcal{O}(n^2)$ to  $\mathcal{O}(n \log n)$ as well as the memory requirement for the elastic operator. This is computationally attractive in conjunction with the use of an iterative solver for the solution of the Jacobian system of the complete non-linear hydromechanical system as discussed later.

In our implementation, the final elasto-static system (\ref{eq:ElasSystem})
is written at all collocation points in the \textbf{local} Cartesian
coordinate system $\bm{e}_{1}=\bm{s}_{1},\bm{e}_{2}=\bm{s}_{2},\bm{e}_{3}=\bm{n}$
of each element. Such a choice  reduces the computational
burden as the constitutive relation of the interface is naturally expressed in such a coordinate system (see section \ref{sec:interface}). The global vector of displacement discontinuity 
is thus ordered as 
\[
\bm{\dd}^T=\left[
\dd_{s_{1}}^{1},
\dd_{s_{2}}^{1}, \dd_{n}^{1},
\cdots,
\dd_{s_{1}}^{j},
\dd_{s_{2}}^{j},\dd_{n}^{j},
\cdots,
\dd_{s_{1}}^{N_{e}}, \dd_{s_{2}}^{N_{e}},\dd_{n}^{N_{e}}
\right]
\]
where $N_{e}$ denotes the total number of collocation points in the mesh which is equal to the number of elements.
The global vector of tractions is ordered similarly: $\bm{T}^T= 
 [ T_{s_{1}}^1,T_{s_{2}}^1,T_{n}^1, ....,T_{s_{1}}^i,T_{s_{2}}^i,T_{n}^i,...,T_{s_{1}}^{N_e},T_{s_{2}}^{N_e},T_{n}^{N_e}]$.
The elastic system (\ref{eq:ElasSystem}) has $N_{m}=n_{d}\times N_{e}$
unknowns where $n_{d}$ denotes the dimension of the problem (2 or
3).

\paragraph{Introducing the effective tractions}

Before discussing the discretization of the fluid-flow inside the discontinuities, we rewrite the boundary element system discretizing the solid balance of momentum by adding and substracting the fluid pore pressure which acts on the normal direction to each element. 
We rewrite the previous equilibrium equation introducing Terzaghi's effective tractions  $\bm{T}^{\prime}=\bm{T} + p \v{n}$. Because the pore pressure acts only in the local normal direction, the vector of effective tractions at the collocation points in the local frame of the boundary element is obtained as 
\[
\bm{T}^{\prime}=\bm{T}+\bm{\mathbb{I}}_{n}\cdot  \bm{p}_{col}
\]
where $ \bm{p}_{col}$ is a vector of size $N_e$ containing the pore-pressure at the location of the collocation points  and $\bm{\mathbb{I}}_{n}$ is a highly sparse matrix of size $\left(n_{d}\times N_{e}\right)\times\left(N_{e}\right)$, ensuring that pore-pressure acts only on the normal component of the local traction, i.e. with ones for the third degree of freedoms of each collocation points.
Adding and substracting pore-pressure in the normal direction of each element, the time-derivative of the discretized elastostatic equilibrium (\ref{eq:ElasSystem}) reduces to:
\begin{equation}
\dot{\bm{T}}^{\prime}-\dot{\bm{T}}^{o}=\mathbb{\bm{E}}\cdot\dot{\bm{\dd}}+\bm{\mathbb{I}}_{n}\cdot\dot{\bm{p}}_{col}-\bm{\mathbb{\eta}}\ddot{\bm{\dd}}	
\label{eq:rate_elastostatic}
\end{equation}
where a dot denotes a time-derivative. In the previous equation, $\dot{\bm{p}}_{col}$ is the rate of pore-pressure at the collocation points. 
%
%

\subsubsection{Elasto-plastic solver}

Assuming for a moment that the rate of change of pore-pressure at the collocation points $\dot{\bm{p}}_{col}$ is known,
we discuss the solution of the elastic system in combination with
an elastoplastic interface law relating the interface effective tractions
and the total displacement discontinuity. Such a non-linear system,
when discretized is akin to a system of differential algebraic equations with inequality constraints. 

We solve in a time-stepping manner from a known solution at time $t^{n}$ to
$t^{n}+\Delta t$ using a classical implicit integration scheme. 
Introducing the increment of total displacement discontinuity, effective tractions and internal variables:
\begin{align*}
\Delta\bm{\dd} & =\bm{\dd}^{n+1}-\bm{\dd}^{n}\\
\Delta\bm{T}^{\prime} & =\bm{T}^{\prime n+1}-\bm{T}^{\prime n}\\
\Delta\bm{\chi}& =\bm{\chi}^{ n+1}-\bm{\chi}^{ n}
\end{align*}
the residuals of the mechanical problem in rate form \eqref{eq:rate_elastostatic} taking the increment of total displacement discontinuity $\Delta\bm{\dd}  $ as the primary unknowns reduce to:
\begin{equation}
\bm{r}_{m}(\Delta\bm{\dd})=\Delta\bm{T}^{\prime, o}-\Delta\bm{T}^{\prime}(\Delta\bm{\dd},\bm{\chi}^{n+1},\bm{T}^{\prime n})+\mathbb{\bm{E}}\cdot\Delta\bm{\dd}+\bm{\mathbb{I}}_{n}\cdot\Delta\bm{p}-\bm{\eta}\left(\Delta\bm{\dd}-\Delta\bm{\dd}^{n}\right)/\Delta t.
\label{eq:res_mech}	
\end{equation}
In this mechanical residual \eqref{eq:res_mech},
we have highlighted the dependence of the increment of the interfaces effective
tractions vector on the increment of displacement discontinuities $\Delta\bm{\dd} $, internal
variables $\bm{\chi}^{n+1}$ and effective tractions at time $t^n$. 
We use a Newton-Raphson scheme to find the root of such 
a non-linear system of equations combined with an elastic predictor - plastic corrector scheme
to integrate locally the interface elasto-plastic law $\Delta\bm{T}^{\prime}(\Delta\bm{\dd},\bm{\chi}^{n+1},\bm{T}^{\prime n},\bm{t})$. 
The numerical integration of  the local interface constitutive relation 
is classical and follow well established algorithms in computational mechanics \citep{SiHu98,deSo11}. The  elastic predictor - plastic corrector scheme is therefore not repeated here, but details are given in Supplemental materials for the constitutive relations used in this work.
The Jacobian of the mechanical problem \eqref{eq:res_mech} $\mathbb{J}_{m}$
is obtained as: 
\begin{equation}
\mathbb{J}_{m}=\mathbb{\bm{E}}-\frac{\bm{\eta}}{\Delta t}-\bm{\mathbb{C}}^{ep}	\text{  with } \bm{\mathbb{C}}^{ep}=\frac{\text{d}\Delta\text{\ensuremath{\bm{T}^{\prime}}}}{\text{d}\Delta\bm{\dd}}
\label{eq:jac_mech}
\end{equation}
 where $ \bm{\mathbb{C}}^{ep}$ is the consistent tangent operator of the interface elasto-plastic law at all collocation points.  It is a block sparse matrix.  
 We combine a convergence criteria on both the residuals vector and on the sequence of $\bm{\bm{x}}^{k}$.
Similarly than for the solution of elasto-plastic initial boundary value problems, the use of the exact  expression of the consistent tangent operator is crucial to achieve robust convergence \citep{SiHu98}. It can be obtained 
by proper analytical linearization of the elastic-predictor scheme, or via automatic code differentiation. 

 
\subsection{Fluid-flow in the discontinuities}	

We discretize the fluid volume balance \eqref{eq:interface_conservation_final} and Darcy's law \eqref{eq:interface_Darcy} along the fractures/faults using a Galerkin finite element method using linear shape function for the pore-pressure. This choice notably allows to naturally handle fractures 
intersection  as the fluid pressure is necessarily continuous between elements. 
The weak form of equations \eqref{eq:interface_conservation_final}-\eqref{eq:interface_Darcy} for the case of an impermeable matrix is given by:
\begin{equation}
\int_{\Gamma}v\frac{\partial \dd_{n}}{\partial t}\text{ d}S+\int_{\Gamma}v\,\frac{w}{M}\frac{\partial p}{\partial t}\text{ d}S+\int_{\Gamma}\bm{\nabla}_{||}v\cdot\left(\frac{w(\v{\dd})k(\v{\dd})}{\mu_{f}}\right)\bm{\nabla}(p-p^o)\text{ d}S=\int_{\Gamma}v\gamma\text{d}S
\label{eq:weak_form_flow}
\end{equation}
where $v$ is a scalar test function with the same regularity than the pore pressure field. We have imposed zero fluid flux at the outer perimeter of the fractures, and assumed a quiescent initial state with an initially hydrostatic pore-pressure $p^o$. 

The fracture mid-plane is discretized with finite elements (segments
in 2D, triangle in 3D) with linear shape functions for simplicity.
The pressure unknowns are thus at the vertex of the mesh. As the displacement discontinuity are discretized with
a piece-wise discontinuous interpolation,  $\dd_{n}$ is constant
over the element. Moreover, the interface permeability is possibly non-linear as function of the normal 
displacement discontinuity $\dd_{n}$  and  as such is assumed constant over the element.
After finite element discretization and assembly,
we obtain the following system of ordinary differential equations: 
\begin{equation}
\mathbb{\bm{V}}\cdot\bm{\dot{\dd}}+\mathbb{\bm{S}}(\bm{\dd})\cdot\dot{\bm{p}}+\mathbb{\bm{L}}(\bm{\dd})\cdot(\bm{p}-\bm{p}^o)=\bm{f}_{f}
\label{eq:flow_discretized}
	\end{equation}
where we have highlighted the dependence of the interfaces permeability and storage coefficient on the displacement discontinuity.
These different finite element matrices at the element level for an element $\Gamma_e$ with $n_n$ nodes are given by:
\begin{eqnarray*} 
   \mathbb{V}^e = \int_{\Gamma_e} \bm{N}^T \cdot \mathbb{I}_{n}^e \text{ d}S,  \qquad  
    \mathbb{S}^e = \int_{\Gamma_e}   \frac{w}{M} \bm{N}^T \cdot\bm{N} \text{ d}S  \\
     \mathbb{L}^e = \int_{\Gamma_e} \frac{w(\dd_n) k(\dd_n)}{\mu_f} \nabla \bm{N}^T \cdot \nabla \bm{N} \text{ d}S, \qquad 
     \bm{f}_{f}^e = \int_{\Gamma_e} \bm{N} \gamma \text{ d}S 
\end{eqnarray*}
 where $\bm{N}$ and $\nabla \bm{N}$ are the matrices of shape functions and their spatial derivatives \citep{ZiTa05a}. $ \mathbb{I}_{en}$ is a $n_n \times n_d$ sparse matrix (where $n_d$ is the problem dimension) such that only the effect of the normal displacement is taken into account: $ (\mathbb{I}^e_n)_{i3}=1$, zero otherwise.  The global finite element matrices for the whole mesh are obtained by classical assembly. We write these global matrices without the superscript $e$.

Like for the mechanical problem, we use an implicit time-integration 
scheme (backward Euler) to obtain the solution at $t^{n+1}=t^{n}+\Delta t$. The residual flow vector for equation \eqref{eq:flow_discretized} is given by:
\begin{equation}
\bm{r}_{f}(\Delta\bm{p},\Delta \bm{\dd})=
\mathbb{\bm{V}}\cdot\Delta\bm{\dd}+\left(\mathbb{\bm{S}}(\bm{\dd}^{n+1})+\Delta t\mathbb{\bm{L}}(\bm{\dd}^{n+1})\right)\cdot\Delta\bm{p}+\Delta t\mathbb{\bm{L}}(\bm{\dd}^{n+1})\cdot(\bm{p}^{n}-\bm{p}^o)-\Delta t\bm{f}_{f}	
\label{eq:res_flow}
\end{equation}

In the case of constant permeability and storage, if the volume change $\Delta\bm{\dd} \cdot \bm{n}$ of the interface is negligible, the flow problem uncouples from mechanics and becomes linear for the pressure increment.  

\subsection{A fully-coupled hydro-mechanical solver}

In the scope of an implicit  time-stepping scheme, the mechanics and flow residuals
are given by eqs \eqref{eq:res_mech}-\eqref{eq:res_flow}. Taking the increment of displacement discontinuity and pressure
as the main unknowns during the step, we obtain 
\begin{align*}
\bm{r}_{m}(\Delta\bm{\dd},\Delta\bm{p}) & =\mathbb{\bm{E}}\cdot\Delta\bm{\dd}-\Delta\bm{T}^{\prime}(\Delta\bm{\dd},\Delta\bm{\chi})+\bm{\mathbb{N}}_{p}\cdot\Delta\bm{p}+\Delta\bm{T}^{o}-\bm{\eta}\left(\Delta\bm{\dd}-\Delta\bm{\dd}^{n}\right)/\Delta t\\ 
\bm{r}_{f}(\Delta\bm{\dd},\Delta\text{\ensuremath{\bm{p}}}) & =\mathbb{\bm{V}}\cdot\Delta\bm{\dd}+\left(\mathbb{\bm{S}}(\Delta\bm{\dd})+\Delta t\mathbb{L}(\Delta\bm{\dd})\right)\cdot\Delta\bm{p}+\Delta t\mathbb{\bm{L}}(\Delta\bm{\dd})\cdot\bm{p}^{n}-\Delta t\bm{f}_{f}
\end{align*}
where the only difference lie in the appearance of matrix $\bm{\mathbb{N}}_{p}$
in the mechanical residuals due to the fact that the pore-pressure
are unknowns at the nodes while the mechanical equations are written
at the collocation points of the boundary element method. This matrix
$\bm{\mathbb{N}}_{p}$ is highly sparse and as before acts only on
the local normal boundary integral equation. For piece-wise constant
displacement discontinuity element, the collocation point is located
at the centroid of the element. As a result the non-zero entries of
$\bm{\mathbb{N}}_{p}$ are chiefly obtained. In fact, 
 using an iterative solver for the solution of the tangent linear system, we do not even need to build this matrix (similarly for $\mathbb{\bm{V}}$).

We find the root of $\mathbf{r}=\left[\begin{array}{c}
\mathbf{r}_{m}\\
\mathbf{r}_{f}
\end{array}\right]$ via a Newton-Raphson scheme - where the unknowns are $\mathbf{X}=\left[\begin{array}{c}
\mathbf{\Delta \dd}\\
\mathbf{\Delta p}
\end{array}\right]$. 
Using the exact consistent tangent operator, we have consistently observed that taking the full Newton step 
results in robust convergence. 
We write the update as 
\[
\mathbf{X}^{k+1}=\mathbf{X}^{k}+ \delta\mathbf{X}^{k}
\]
 where $\delta\mathbf{X}^{k}$ is the descent direction, solution of the linearized Jacobian system 
\[
\mathbf{r}(\mathbf{X}^{k})+(\mathbf{\nabla_{X}}\mathbf{r}(\mathbf{X}^{k}))\cdot\delta\mathbf{X}^{k}=\bm{0}
\] 

\subsubsection{Solution of the Jacobian system}
\label{section_sol_jac_system}

Using the notation of the previous sections, the Jacobian matrix of
the coupled hydro-mechanical system has a block structure given by 
\[
\mathbf{\nabla_{X}}\mathbf{r}(\mathbf{X}^{k})=\left[\begin{array}{cc}
\mathbb{E}-\frac{\bm{\eta}}{\Delta t}-\bm{\mathbb{C}}^{ep} & \mathbb{N}_{p}\\
\mathbb{V}+\left(\mathbf{\nabla}_{\Delta \dd}\mathbb{S}+\Delta t\mathbf{\nabla}_{\Delta \dd}\mathbb{L}\right)\cdot\mathbf{\Delta p}^{k} & \mathbb{S}+\Delta t\mathbb{L}+\Delta t\left(\mathbf{\nabla}_{\Delta p}\mathbb{L}\right)\cdot\mathbf{\Delta p}^{k}
\end{array}\right]
\]
where $\Delta t\left(\mathbf{\nabla}_{\Delta p}\mathbb{L}\right)\cdot\mathbf{\Delta p}^{k}$
and $\left(\mathbf{\nabla}_{\Delta \dd}\mathbb{S}+\Delta t\mathbf{\nabla}_{\Delta \dd}\mathbb{L}\right)\cdot\mathbf{\Delta p}^{k}$
are consistent tangent operator for the flow problems, while $\bm{\mathbb{C}}^{ep}$
is the elasto-plastic consistent tangent operator of the interface.

The block of the Jacobian corresponding to the flow parts are obtained via differentiation of the corresponding part of the residual at the element level.
Notably 
\[
\mathbf{\nabla}_{\Delta \dd}\mathbb{S}\cdot\mathbf{\Delta p}
\]
 at the element level, for piece-wise constant displacement discontinuity element, ends up being a diagonal matrix (separating the terms coming from $\dd_n/M$ (i.e with a potential non-linearity of $M$):
\[
\left(\int_{\Gamma_{e}}\bm{N}^{T}\,\frac{1}{M}\,\bm{N}\text{ d}S\right)\cdot\mathbf{\Delta p_{e}}+w\left(\int_{\Gamma_{e}}\bm{N}^{T}\,\frac{\partial (1/M)}{\partial \dd_n}\:\bm{N}\text{ d}S\right)\cdot\mathbf{\Delta p_{e}}
\]
where $w=w_o+\dd_n$  is the current thickness of element $e$.
Similarly 
\[
\mathbf{\nabla}_{\Delta \dd}\mathbb{L}\cdot\mathbf{\Delta p}
\]
 is at the element level 
\[
\left(\int_{\Gamma_{e}} \frac{1}{\mu_f}\frac{\partial (w \times k)}{\partial\Delta \dd_{n}} \bm{\nabla}\bm{N}^{T}\cdot\bm{\nabla}\bm{N}\text{ d}S\right)\cdot\Delta p_{e}.
\]

We can schematically rewrite the Jacobian matrix in 4 separate blocks as:
\[
\left[\begin{array}{cc}
{\bf A}_{11} & {\bf A}_{12}\\
{\bf A}_{21} & {\bf A}_{22}
\end{array}\right]\left[\begin{array}{c}
{\bf x}_{1}\\
{\bf x}_{2}
\end{array}\right]=\left[\begin{array}{c}
{\bf b}_{1}\\
{\bf b}_{2}
\end{array}\right]
\]
with 
\begin{align*}
{\bf A}_{11} & =\mathbb{E}-\frac{\bm{\eta}}{\Delta t}-\bm{\mathbb{C}}^{ep},\qquad{\bf A}_{12}=\mathbb{N}_{p}\\
{\bf A}_{21}= & \mathbb{V}+\left(\mathbf{\nabla}_{\Delta \dd}\mathbb{S}+\Delta t\mathbf{\nabla}_{\Delta \dd}\mathbb{L}\right)\cdot\mathbf{\Delta p}^{k}\\
{\bf A}_{22}= & \mathbb{S}+\Delta t\mathbb{L}+\Delta t\left(\mathbf{\nabla}_{\Delta p}\mathbb{L}\right)\cdot\mathbf{\Delta p}^{k}
\end{align*}
we note that the flow blocks need to be updated at every iteration
of the Newton-Raphson (as $\mathbb{L},\,\mathbb{S}$ depends
on opening and possibly pressure). 
In order to benefit from the hierarchical representation of the collocation boundary element matrix, we solve this Jacobian system of equations via an iterative method using a triangular block pre-conditioner. 
Such a preconditioning approach is very similar to the ones used for the solution of hydro-mechanical problems via finite element (see for examples \cite{Prev97,FeBe10,WhCa16} among many others, as well as the recent work of \cite{ZaBe26} in relation to fracture contact within a finite volume spatial discretization).

Specifically, we use an upper block triangular pre-conditioner,
\[
\mathbf{P}_{up}=\left[\begin{array}{cc}
{\bf A}_{11} & {\bf A}_{12}\\
{\bf 0} & \tilde{{\bf S}}
\end{array}\right]\qquad\tilde{{\bf S}}={\bf A}_{22}-{\bf A}_{21}{\bf \tilde{A}}_{11}^{-1}{\bf A}_{12}
\]
where $\tilde{{\bf S}}$ is an approximation of the Schur complement
and ${\bf \tilde{A}}_{11}$ an approximation of ${\bf A}_{11}$ for
which we can "cheaply" compute an inverse .
We then transform the original system 
\[
\mathbf{A}\cdot\mathbf{x}=\mathbf{b}
\]
into 
\[
\left(\mathbf{A}\mathbf{P}_{up}^{-1}\right)\hat{\mathbf{x}}=\mathbf{b},~\text{and}~\mathbf{P}_{up}\mathbf{x}=\hat{\mathbf{x}},
\]
where
\[
\mathbf{P}_{up}^{-1}=\left[\begin{array}{cc}
{\bf A}_{11}^{-1} & -{\bf A}_{11}^{-1}{\bf A}_{12}\tilde{{\bf S}}^{-1}\\
{\bf 0} & \tilde{{\bf S}}^{-1}
\end{array}\right]
\]
The above system of equations is first solved for $\hat{\mathbf{x}}$ by solving $\left(\mathbf{A}\mathbf{P}_{up}^{-1}\right)\hat{\mathbf{x}}=\mathbf{b}$ using an iterative solver such as BICGSTAB or GMRES.
 The application of the preconditioner in the above computation is akin to solving $\mathbf{P}_{up}\mathbf{z}=\mathbf{y}$,
which we can rewrite as 
\begin{align*}
{\bf z}_{2} & ={\bf \tilde{{\bf S}}}^{-1}\cdot{\bf y}_{2}\\
{\bf A}_{11}{\bf z}_{1} & ={\bf y}_{1}-{\bf A}_{12}{\bf z}_{2}
\end{align*}
We therefore see that at each iteration of the iterative tangent solver,
we need to solve a mechanical system (involving ${\bf A}_{11}$). 
Our experience has shown that, 
unfortunately, using an approximation
for ${\bf A}_{11}$ at the pre-conditioner level (for example a diagonal
or an ILU decomposition of the near-diagonal terms) is not robust when
large non-linearities associated with fracture opening occurs. We therefore solve the ${\bf A}_{11}$ system using
an iterative method such as BICGSTAB with a Jacobi pre-conditioner. 
The solution of the overall linear system requires the use of nested iterative solvers. 
 Once we have calculated $\hat{\mathbf{x}}$ using the above, we compute $\bm{\mathbf{x}}$ by solving the $~\mathbf{P}_{up}\mathbf{x}=\hat{\mathbf{x}}$ which mimics the same steps as applying the preconditioner as explained above.

The solver makes use of two possible approximation for ${\bf A}_{11}^{-1}$, used as part of the Schur complement approximation ${\bf \tilde{{\bf S}}}$ as well as for preconditioning when solving on ${\bf A}_{11}$. The first and simplest one is to take $\tilde {\bf A}_{11} = \text{diag}({\bf A}_{11})$. Under this choice ${\bf \tilde{{\bf S}}}$ can be expressed explicitly as a sparse matrix and the application of ${\bf \tilde{{\bf S}}}^{-1}$ is performed by means of an ILU-T factorization \citep{Li2005}. This simple choice proved to be sub-optimal, leading to a poor preconditioning of the numerous solve operations on ${\bf A}_{11}$ as well as a poor quality of the Jacobian preconditioner as the number of yielded elements grew. To improve this the second option is to define $\tilde {{\bf A}_{11}}$ as a sparse matrix containing a subset of the information of ${\bf A}_{11}$. We recall that ${\bf A}_{11} = \mathbb E + \mathbb C^{ep}$, with $\mathbb C^{ep}$ the consistent tangent operator, a block diagonal matrix, and $\mathbb E$ the hierarchical matrix associated to the BEM. We define $\tilde {{\bf A}}_{11}$ as the sum of $\mathbb C^{ep}$ and a fraction $p \in ]0,1]$ of the entries with highest absolute value among the entries of the full-rank blocks (near-field terms) of the hierarchical matrix $\mathbb E$. The application of ${\bf A}_{11}^{-1}$ is then done by means of an ILU factorization. Under this definition of ${\bf A}_{11}^{-1}$, ${\bf \tilde{{\bf S}}}$ is no longer a sparse matrix and we rely on iterative solvers for the application of ${\bf \tilde{{\bf S}}}^{-1}$. We use the previously defined first version of ${\bf \tilde{{\bf S}}}^{-1}$, obtained with $\tilde {\bf A}_{11} = \text{diag}({\bf A}_{11})$, as preconditioner. 
This strategy allows to greatly improve the quality of the ${\bf A}_{11}$ solver preconditioning as well as the quality of the overall Jacobian block preconditioner. 
We present the performance of this two possible approximation in sub-section \ref{subsec:computational_insights}. 
The choice of the value for $p$ is a trade-off between the speed-up of linear solvers application and the cost of computing the ILU factorization. We use $p=0.01$ by default.

\subsubsection{Scaling of the unknowns and residuals}

To reduce floating point cancellation, we apply scaling factors to the mechanical and flow unknowns and residuals. These scaling factors are aiming at making these 4 quantities $\mathcal O(1)$. As these change by orders of magnitude during a simulation, the scaling factors need to be adapted. We define how they are updated for the mechanical unknowns and residuals, the procedure is identical for the flow subproblem. We note $s_X$ the scaling factor applied to the unknown vector, the displacement discontinuity $\Delta {\bm \delta}$, and $s_R$ the scaling factor applied to the mechanical residuals. They have units of m$^{-1}$ and Pa$^{-1}$ respectively. We use scaling values on the rate of displacement discontinuity and on the residuals taken as the maximum absolute values of the associated vectors at the previous time step $n$:
\begin{equation}
    L_{\dot X} = \max(\dot{\Delta {\bm \delta}_n}), \qquad {L}_R = \max({\bm r}_{n}),
\end{equation}

with $\dot{\Delta {\bm \delta}_n} = \Delta {\bm \delta}_n / \Delta t_n$. For stability, these two scales are smoothed in time using the previous values $ L_{\dot X}^{\text{old}}$ and ${L}_R^{\text{old}}$ using a smoothing parameter $\alpha \in [0,1]$:

\begin{equation}
    L_{\dot X} \leftarrow \alpha \, L_{\dot X}^{\text{old}} + (1-\alpha)\,  L_{\dot X}, \qquad
    {L}_R \leftarrow \alpha \, {L}_R^{\text{old}} + (1-\alpha)\, {L}_R.
\end{equation}
To prevent excessively large or small updates, the ratio of the new to the old
scaling factor is clipped to the interval $[\rho_{\min}, \rho_{\max}]$:
\begin{equation}
    L_{\dot X} \leftarrow L_{\dot X}^{\text{old}} \cdot
    \operatorname{clip}\!\left(\frac{L_{\dot X}}{L_{\dot X}^{\text{old}}},\,
    \rho_{\min},\, \rho_{\max}\right),
\end{equation}
and analogously for $L_R$, where $\operatorname{clip}(x, a, b) = \max(a, \min(x, b))$. The three controlling parameters ($\alpha, \rho_{\min},\,\rho_{\max}$) have been chosen from a series of numerical simulations but the overall algorithm is robust to their choice. We use  $\alpha = 0.62$, $\rho_{\min} = 0.84$ 
and $\rho_{\min} = 1.16$ throughout.

Finally we obtain the scaling factors as:

\begin{equation}
    s_X = \frac{1}{L_{\dot X} \Delta t_{n+1}}, \qquad
    s_R = \frac{1}{{L}_R}
\end{equation}

\subsection{Adaptive time-stepping}
 
Adaptive time-stepping for a system of ODEs requires an estimation of
the local truncation error (LTE). This is not directly available for
a fully implicit backward-Euler scheme. One can estimate such an LTE in a crude way
by comparing a simple ``explicit'' prediction from the time-derivative
estimated at the previous step and the solution of the non-linear
implicit step (see \cite{ShSl02,Dier13} for details). 
Denoting $\bm{X}^{n}$ and $\bm{\dot{X}}^{n}$ the solution and rate
at time $t_{n},$ and similarly $\bm{X}^{n}$ and $\bm{\dot{X}}^{n}$
at $t_{n}=$ $t_{n-1}+\Delta t_{n}$ . A first-order explicit extrapolation
provides 
\[
\tilde{\bm{X}}_{n}=\bm{X}_{n-1}+\Delta t_{n}\bm{\dot{X}}_{n-1}
\]
 and a second-order accurate one
\[
\bm{X}_{n}=\bm{X}_{n-1}+\Delta t_{n}\bm{\dot{X}}_{n-1}+\frac{1}{2}\Delta t_{n}^{2}\bm{A}
\]
where the acceleration can be estimated as 
\[
\bm{A}=\frac{\bm{\dot{X}}_{n}-\bm{\dot{X}}_{n-1}}{\Delta t_{n}}.
\]
The LTE is estimated as time $t_{n+1}$ as the norm of the difference
between the first order and second order estimate \citep{KaBi02}:
\[
LTE(t_{n})\approx e_{n}=\frac{\Delta t_{n}}{2}\|\bm{\dot{X}}_{n}-\bm{\dot{X}}_{n-1}\|.
\]
 It is worth recalling that the implicit scheme previously discussed
solve for $\Delta\bm{X}$, in other word for the rate of change of
the solution at $t_{n+1}$.
More specifically, to ensure a better scaling, we estimate one LTE
for pressure and one for displacement discontinuity and scale it with
the norm of the current solution:
\[
e_{n+1}(\bm{p})=\frac{\Delta t_{n}}{2}\|\bm{\dot{p}}^{n+1}-\bm{\dot{p}}^{n}\|/\|\bm{p}^{n+1}\|\qquad e_{n+1}(\bm{d})=\frac{\Delta t_{n}}{2}\|\bm{\dot{d}}^{n+1}-\bm{\dot{d}}^{n}\|/\|\bm{d}^{n+1}\|
\]
and takes the maximum of these two estimates to adapt the time step. 

With such an LTE in hand, we use a PID controller similar to the one described in \citet{Sode02,SoWa06}
to adapt the time-step size for the subsequent step. Denoting $\Delta t_{n+1}$,
this new time-step size is estimated as:
\[
\Delta t_{n+1}=\Delta t_{n}
\times \left(\frac{\Delta t_{n}}{\Delta t_{n-1}} \right)^{-1/4}
\times\left(\frac{\epsilon_{TOL}}{e_{n+1}}\right)^{1/4}\times\left(\frac{e_{n}}{\epsilon_{TOL}}\right)^{-1/4}
\times \left( \frac{n_{its}}{n_{target}} \right)^{-1/3}
\]
with $\epsilon_{TOL}$ the target error, $n_{its}$ the number of Newton-Raphson iterations of the previous time-step, and  $n_{target}$ a target iteration count (typically 6).
In addition, we enforce a minimum time-step size, and enforce a maximum factor $M$ such that at most $\Delta t_{n+1}=M\Delta t_{n}$. We typically use $M=1.05$ to $M=1.2$. 
To ensure $e_{n+1}\ne0$, we add the square root of machine precision to the estimate of $e_{n+1}$.

%
\section{A series of verification examples for propagating fluid-driven ruptures}

We now turn to the verification of the accuracy of the presented numerical scheme on a series of four fluid-driven rupture propagation problems whose solution is now available. 
These tests are a must-pass for any solver. More importantly, the comparisons must be performed for sufficiently long propagation distance (/injection duration) 
to demonstrate relevance at engineering scales.

All the simulations are performed in 3D, even though the available analytical solutions are all for an axisymmetric configuration. We use for the convergence of the Newton solver, an absolute tolerance of $10^{-3} \times \|\bm{T}^o\|$ (where $ \|\bm{T}^o\|$ is the L2 norm of the initial tractions over the whole mesh and a relative tolerance of $10^{-3}$ for the increment of the solution vectors (increment of displacement discontinutiy and pore-pressure). 
 In addition, we use a goal $\epsilon_{TOL}=10^{-3}$ for automatic time-step adaptation.
 The details of all the parameters and discretization used are reported in Supplemental Materials for all the problem tests discussed here. 

\begin{figure}
    \centering
    \includegraphics[width=0.5\linewidth]{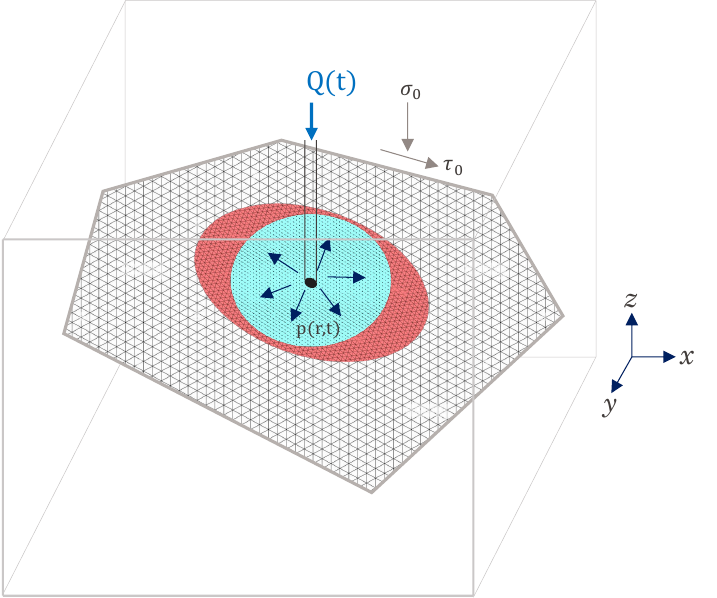}
    \caption{Schematic of the 3D problem setup. A polygonal fault is embedded in a three-dimensional domain (x,y,z) and subjected to far-field in-situ tractions, with $\sigma_0$ denoting the compressive normal stress and $\tau_0$ the initial shear traction acting on the fault plane. Fluid is injected through a line source at a prescribed rate Q(t), generating a transient pressure field p(r,t) that diffuses along the fault and may promote shear slip. Red patch denotes the portion of the fault that has yielded/slipped.}
    \label{fig:domain}
\end{figure}

\subsection{Frictional ruptures without permeability changes}

\subsubsection{The constant friction case}
    
For the case of an impermeable homogeneous medium and a planar interface with homogeneous and constant hydraulic properties, it is possible to obtain an analytical solution for the propagation of a circular frictional rupture driven by a constant injection rate from a point-source - for the evolution of the rupture front and slip profile \citep{SaLe22,Vies24}. 
For a constant and uniform friction coefficient (and initial stress), the rupture radius $R(t)$ evolves in a self-similar way, proportional to the pore-pressure diffusion front $\sqrt{4 \alpha t}$ (where $\alpha = k M/(\mu)$ the hydraulic diffusivity of the interface):
\[
R(t)=\lambda(\mathcal{T}) \sqrt{4 \alpha t}
\]
$\lambda(\mathcal{T})$ is an amplification factor solely function of the fault-injection stress dimensionless parameter 
\begin{equation}
\mathcal{T} =\frac{f\,\sigma^\prime _o-\tau_o}{f \Delta p_*},     \qquad \Delta p_* =\frac{Q \mu }{4 \pi k w } 
\label{eq:T_definition}
\end{equation}
where $Q$ is the injection rate, $\mu$ the fluid viscosity, $k$ the intrinsic permeability, and $w$ the hydraulic aperture. The parameter \( \mathcal{T} \) quantifies the relative distance to failure compared to the strength of injection, and governs the rupture dynamics. Based on the value of T, the fluid-driven rupture 
has distinct characteristics. For  $\mathcal{T} \ll 1$, the rupture front is ahead of the pore-pressure front ($\lambda>1$) - the rupture propagates in the so-called critically stressed regime. For large value $\mathcal{T} > 1$, the opposite occurs: the rupture radius remains within the pressurized region ($\lambda<1$) -  marginally pressurized regime. The solutions obtained in \citep{SaLe22,Vies24} for a circular rupture are valid for a zero Poisson's ratio $(\nu = 0)$. For $\nu \ne 0$, the rupture elongates as an ellipse (as function of both $\nu$ and $\mathcal{T}$), but the rupture area is strictly equal to the circular  $(\nu = 0)$ case - as shown in \cite{SaLe22}.  

We present a verification tests for the case of a critically stressed fault having $\mathcal{T} = 0.05 $ for circular rupture $(\nu = 0)$ (a simulation for a marginally pressurized case $\mathcal{T}=4$ is reported in Supplemental Materials). 
We also present a series of simulation for $\mathcal{T} = 0.05 $ with increasing value of $\nu$, and compare with the results presented in \cite{SaLe22}.
 We model the fault as a hexagonal plane, $\Gamma$ with major axis spanning 1000 m, located along $z=0$. The fluid is injected at a point source at the center of the fault.
The complete set of material, in-situ and injection parameters are listed in supplemental.

\begin{figure}[H] 
    \centering
    \begin{subfigure}[b]{0.48\linewidth}
        \centering
        \includegraphics[width=\linewidth]{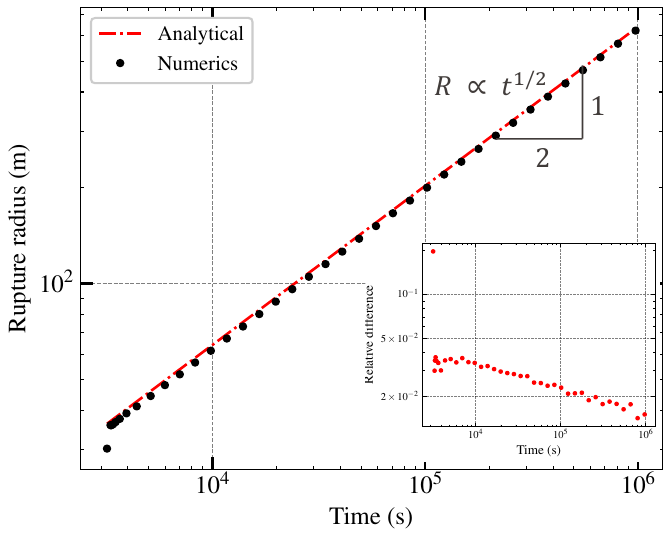}
        \caption{}
        \label{fig:lam_plot}
    \end{subfigure}%
    \hfill
    \begin{subfigure}[b]{0.48\linewidth}
        \centering
        \includegraphics[width=\linewidth]{ 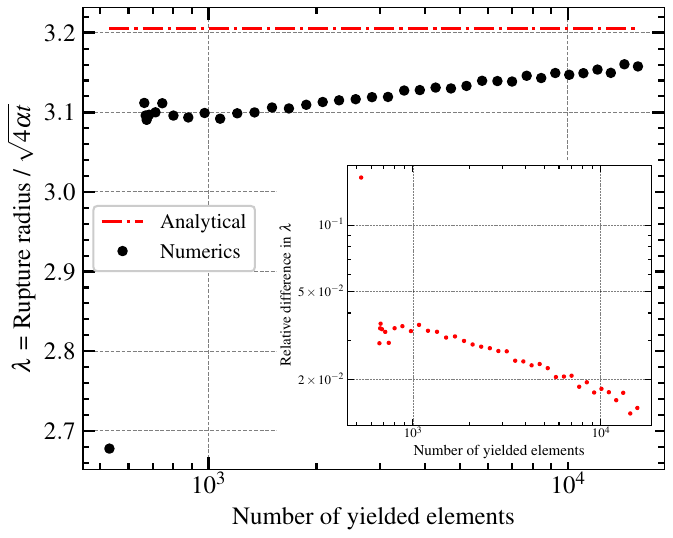}
        \caption{}
        \label{fig:err_lam}
    \end{subfigure}
    \caption{(a) Evolution of the rupture radius with time obtained from the 3D numerical results and analytical solution for a circular rupture for a critically stressed case ($\mathcal{T} = 0.05$).
    b) Evolution of the amplification factor $\lambda = R(t)/\sqrt{4\alpha t}$  as function of the number of yielded elements in the mesh. 
     Relative errors in inset, plots in log-log or log-linear.
     }
    \label{fig:rupture_overview}.
\end{figure}
 
The accuracy of the numerical scheme can be first assessed by examining the evolution of the rupture radius with time reported on Figure~\ref{fig:rupture_overview}(a). At each time-step, the rupture radius is extracted from the 3D numerical results by averaging the distance between the injection point and the furthest away yielded elements for all azimuthal angles (from the injection point). 

 The numerical results closely match the analytical solution, confirming that the solver accurately captures the self-similar propagation of the rupture front. A different quantitative measure of this agreement is provided in Figure~\ref{fig:rupture_overview}(b), where the relative error in the amplification factor $\lambda=R/\sqrt{4 \alpha t}$ is plotted as a function of the number of yielded elements. 
Because the numerical solver is based on fix mesh, the rupture will always be initially poorly resolved as only a few elements exhibit plastic slip. 
The error decreases monotonically as the rupture expands and is resolved by a larger number of elements, demonstrating proper convergence of the rupture propagation dynamics - which here follows a sub-linear convergence associated with the piece-wise 
constant displacement discontinuity collocation boundary element method used \citep{RyNa85}.

\begin{figure}[]
    \centering
    \begin{subfigure}[b]{0.5\linewidth}
        \centering
        \includegraphics[width=\linewidth]{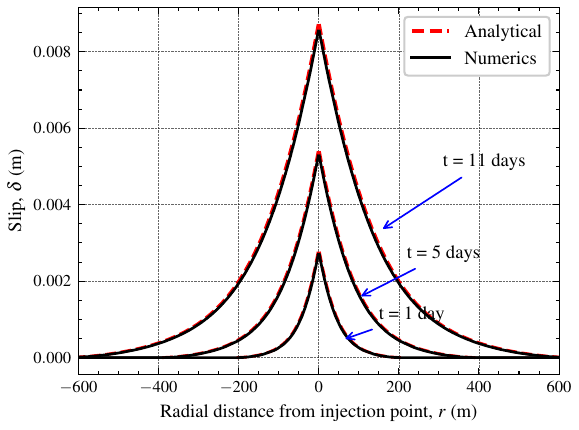}
        \caption{}
        \label{fig:slip_profile_cs}
    \end{subfigure}%
    \hfill
    \begin{subfigure}[b]{0.46\linewidth}
        \centering
        \includegraphics[width=\linewidth]{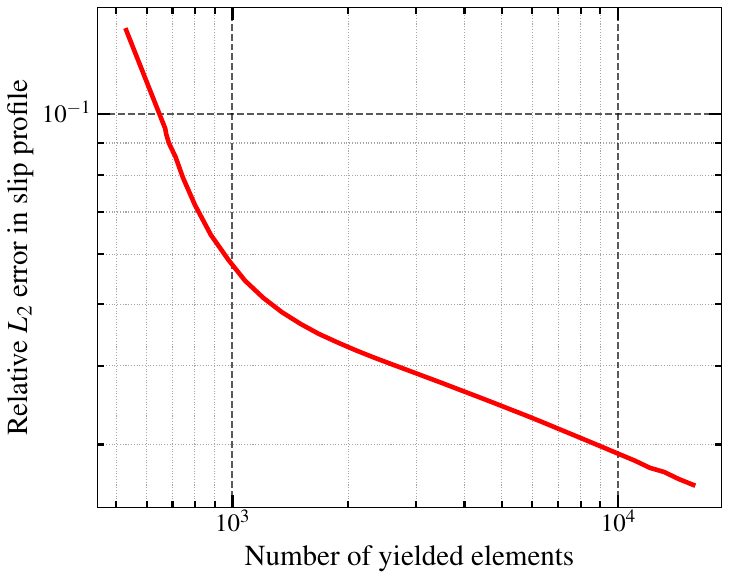}
        \caption{}
        \label{fig:err_slip}
    \end{subfigure}
    \caption{(a) Comparison between analytical and numerical slip profiles (along a radial direction) at three different times ($t = 1, 5,$ and $11$ days) for a circular rupture under critically stressed conditions ($\mathcal{T} = 0.05$). 
    (b) Relative $L_2$ error in the slip profile as a function of the number of yielded elements.}
    \label{fig:slip_overview}
\end{figure}

Figure~\ref{fig:slip_overview}(a) presents a comparison between the analytical solution derived by \cite{Vies25} and the numerical slip profiles (along a radial direction)
at three representative times during the injection process. The numerical solution accurately reproduces the analytical self-similar slip distribution, with excellent agreement observed both near the injection point and toward the rupture front. 
The corresponding relative $L_2$ error in the slip profile as a function of the number of yielded elements is presented in Figure~\ref{fig:slip_overview}(b). Similarly than for the rupture radius, the error decreases monotonically as the rupture grows and more elements falls within the slipping region. This behavior reflects the progressive refinement of the effective resolution of the slipping region, and confirms the convergence of the numerical solution toward the analytical benchmark.

\paragraph{Non-circular ruptures \\}

\begin{figure}[]
    \centering
    \begin{subfigure}[b]{0.50\linewidth}
        \centering
        \includegraphics[width=\linewidth]{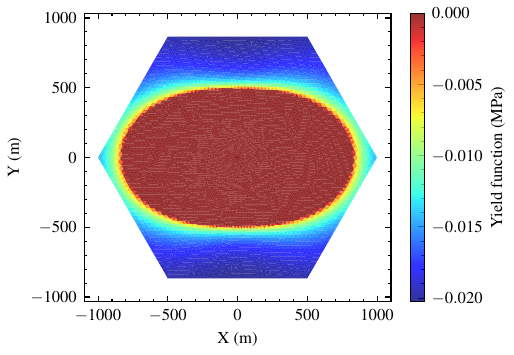}
        \caption{}
        \label{fig:yf_noncircular}
    \end{subfigure}%
    \hfill
    \begin{subfigure}[b]{0.44\linewidth}
        \centering
        \includegraphics[width=\linewidth]{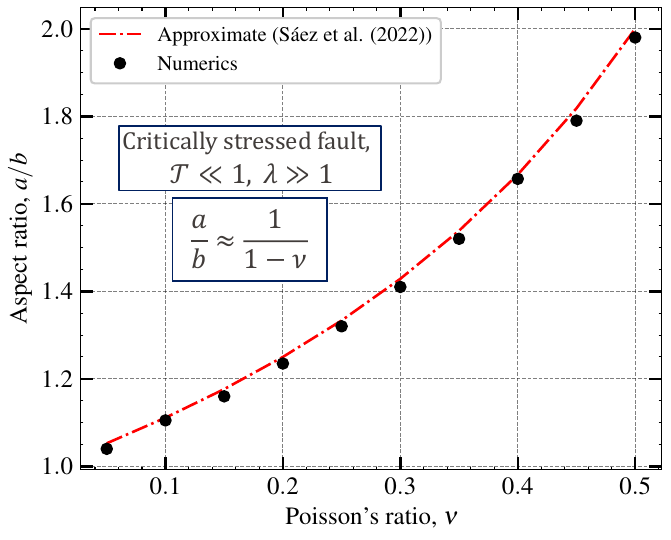}
        \caption{}
        \label{fig:aspect_ratio_cs}
    \end{subfigure}
    \caption{
    Non-circular rupture in a critically stressed fault with $\mathcal{T}=0.05$.
    (a) Spatial distribution of the yield function after 11 days of injection for $\nu=0.45$, displaying an elliptical rupture front. 
    (b) Comparison between the numerical and analytical rupture aspect ratio $a/b$ as a function of Poisson's ratio $\nu$. The approximate analytical estimate is taken from \citet{SaLe22}.
    }
    \label{fig:noncircular_verification}
\end{figure}

We illustrate the influence of Poisson’s ratio on the rupture geometry. Non-zero Poisson's ratio leads to deviations from circular rupture fronts \citep{SaLe22}. Figure~\ref{fig:noncircular_verification}(a) illustrates the spatial distribution of the yield function for a critically stressed fault ($\mathcal{T} = 0.05$) with $\nu = 0.45$ after 11 days of injection. In contrast to the circular case, the rupture front exhibits a pronounced elongation which can be well approximated by an elliptical shape characterized by a major axis $'a'$ and a minor axis $'b'$. 
\cite{SaLe22} have proposed an approximation for the rupture aspect ratio $a/b$ as a function of Poisson’s ratio in both the critically stressed and marginally pressurized  limits. 

Figure~\ref{fig:noncircular_verification}(b) display the numerical evaluation of the rupture aspect ratio as function of  Poisson’s ratios for the critically stressed case ($\mathcal{T} = 0.05$). We extract the aspect ratio from the numerical results by fitting an ellipse to the rupture radius extracted at each time step. The numerical results are in excellent agreement with the approximate prediction proposed in \cite{SaLe22}. This test confirms that the proposed numerical scheme accurately captures the influence of 3D elastic coupling on rupture geometry.

\subsubsection{Slip-weakening friction cases}

We now consider the case of a planar fault governed by a linear slip-weakening friction law. In contrast to the constant-friction cases discussed previously, the friction coefficient is no longer constant but decreases from a peak value $f_p$ to a residual value $f_r$ over a characteristic slip-weakening distance $d_c$. This additional weakening mechanism introduces the possibility of a transition from quasi-static to dynamic slip, depending on the initial stress state, the amount of frictional weakening, and the magnitude of the injection-induced pressurization.
Following \citet{SaLe24}, the response of a circular fluid-driven rupture ($\nu=0$) with slip-weakening friction can be characterized by three dimensionless parameters: i) the pre-stress ratio $ \mathcal{S} = \tau_o/f_p \sigma^\prime_o$, the ratio of residual to peak friction $ \mathcal{F} = f_r/f_p$, and the ratio of the characteristic injection over-pressure (previously defined for the constant friction case) over the initial normal effective stress $  \mathcal{P} = \Delta p_*/\sigma^\prime_0$. Depending on the values of these parameters, \citet{GaGe12,SaLe24} identified different regimes of rupture propagation. Notably, an unabated dynamic rupture always nucleate when the residual fault strength is below the initial shear stress $(\mathcal{S}>\mathcal{F})$.
In the unconditionally stable domain $(\mathcal{S}<\mathcal{F})$, the rupture either remain fully quasi-static, or may experience a transient dynamic episode that subsequently arrests before quasi-static propagation resumes. 

We present here two  3D simulations of injection at constant rate  in the unconditionally stable domain, 
using the same value of the pre-stress ratio and residual-to-peak friction ratio,
\begin{equation}
    \mathcal{S}=0.6, \qquad \mathcal{F}=0.7,
\end{equation}
and varying the dimensionless overpressure ratio $\mathcal{P}$.
The first case, with $\mathcal{P}=0.05$, corresponds to regime (denoted R1 in \cite{SaLe24}) where the rupture always remain quasi-static although it accelerates as the weakening to residual friction localize in a process zone at the tip of the circular rupture. 
The second case, with $\mathcal{P}=0.035$, exhibit a transient instability associated with frictional weakening (R2 in \cite{SaLe24}). The arrest of the instability is associated with the frictional rupture catching up with fluid pressure front
(see \cite{SaLe24} for detailed discussions). 
These tests are important to
verify that the quasi-dynamic implementation of our solver correctly captures both the stable quasi-static growth and the transient instability associated with slip weakening.

At early time, before significant frictional weakening has developed over the slipping region, the rupture is expected to follow the constant-friction similarity solution obtained by taking the friction coefficient equal to its peak value $f_p$.  
At late time, once the weakening zone has localized in a small process zone (compared to the rupture size) close to the rupture front, the rupture growth can be well approximated by a front-localized energy balance: $ G=G_c$
where $G$ is the energy release rate and $G_c$ is the fracture energy associated with the slip-weakening law. We refer to \cite{SaLe24} for details. We therefore have two analytical solutions (valid respectively at early and late times) to verify our numerical results for the evolution of the rupture radius. 

\begin{figure}[] 
    \centering
    \begin{subfigure}[b]{0.48\linewidth}
        \centering
        \includegraphics[width=\linewidth]{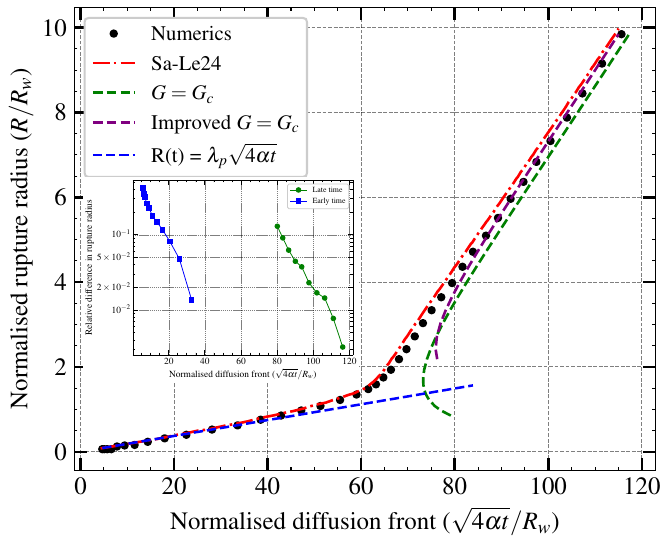}
        \caption{}
        \label{fig:sw_3d_r1}
    \end{subfigure}
    \hfill
    \begin{subfigure}[b]{0.48\linewidth}
        \centering
        \includegraphics[width=\linewidth]{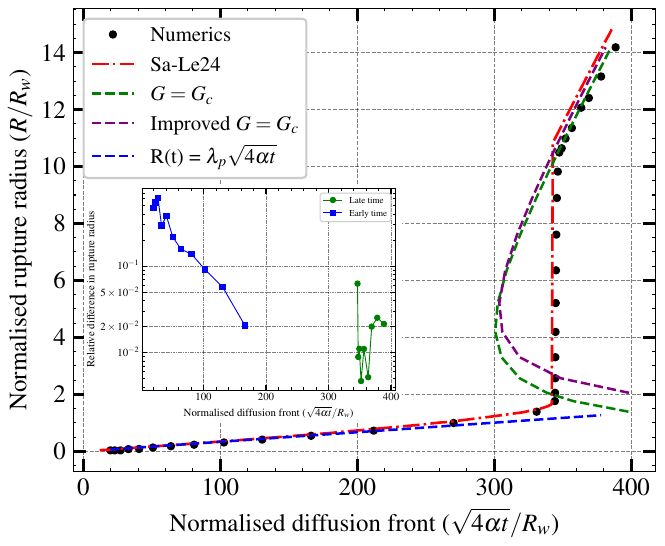}
        \caption{}
        \label{fig:sw_3d_r2}
    \end{subfigure}
    \caption{3D circular fluid-driven rupture with linear slip-weakening friction. Evolution of the normalized rupture radius $R/R_w$ as a function of the normalized diffusion front $\sqrt{4\alpha t}/R_w$ for $\mathcal{S}=0.6$ and $\mathcal{F}=0.7$. 
    (a) $\mathcal{P}=0.05$: the rupture remains quasi-static during the whole propagation. 
    (b) $\mathcal{P}=0.035$: the rupture first propagates quasi-statically, nucleates a transient dynamic rupture, arrests, and subsequently resumes quasi-static propagation. 
    Black dots denote the numerical results. The red dashed line shows the reference solution of \citet{SaLe24} obtained with an axisymmetric solver. The blue dash-dotted line corresponds to the early-time constant-friction similarity solution evaluated with the peak friction coefficient, $R(t)=\lambda_p\sqrt{4\alpha t}$. The green dashed line shows the rupture radius predicted by the front-localized energy-balance approximation, $G=G_c$, which is valid once the process zone is localized. The purple line shows the improved energy-balance solution accounting for the finite size of the process zone (after \cite{GaGe12}). Insets show the relative errors with respect to the early-time constant-friction solution and the late-time energy-balance solution.}
    \label{fig:sw_3d_r1_r2}
\end{figure}

The evolution of the rupture radius for these two cases is displayed in Fig.~\ref{fig:sw_3d_r1_r2}. 
For $\mathcal{P}=0.05$, the rupture remains fully quasi-static. The numerical rupture radius initially follows the constant-friction solution based on $f_p$. This early-time agreement is also quantified by the relative error shown in the inset of Fig.~\ref{fig:sw_3d_r1_r2}a. As slip accumulates and the friction coefficient decreases behind the rupture front, the numerical solution progressively departs from the peak-friction similarity solution. The rupture subsequently accelerates and tends toward the solution obtained from the front-localized energy balance. The late-time relative error (with respect to the front-localized energy balance) shown in the inset decreases as the rupture radius becomes large compared to the slip-weakening process zone scale, indicating the progressive validity of the front-localized energy balance approximation.

For $\mathcal{P}=0.035$ (Fig.~\ref{fig:sw_3d_r1_r2}b), the numerical solution is again well approximated by the constant-friction solution during the first stage of propagation, as long as the frictional weakening remains limited. In this case, the rupture nucleates a transient dynamic episode, which subsequently arrests before quasi-static propagation resumes. The late-time evolution is then again well captured by the front-localized energy balance. The comparison with the energy-balance solution shows that the solver reproduces not only the quasi-static propagation phases, but also the transition through a transient dynamic instability and its subsequent arrest. The time of nucleation of the dynamic episode is also matching well with the results of \cite{SaLe24} obtained using an axisymmetric solver. We observe that the 3D numerical results appear to be more accurate than the results of  \cite{SaLe24} at late time. This may be associated with the use of ring constant displacement discontinuity element in \cite{SaLe24}.

It is important to note that, in both simulations, the early-time and late-time comparisons test different aspects of the numerical implementation. The early-time agreement with the constant-friction solution verifies the response before significant weakening develops. The late-time agreement with the front-localized energy balance verifies the correct release of fracture energy once the weakening process zone is localized near the rupture front. The  differences between the numerical results and the energy-balance solution are expected at intermediate times, since the front energy balance balance assumes a localized process zone whereas the numerical model resolves a finite slip-weakening zone. The decrease of the relative error with increasing rupture radius is consistent with convergence toward the small-scale yielding limit.

This example verifies the ability of the quasi-dynamic solver to reproduce the different stages of a fluid-driven slip-weakening rupture, from the early Coulomb-like propagation controlled by the peak friction coefficient to the late propagation controlled by a front-localized fracture energy balance. It also verifies that the solver captures transient dynamic rupture and arrest accurately.

\subsection{Dilatant frictional ruptures with permeability changes}

\cite{Dunh24} has derived a solution for the growth of a dilatant circular ($\nu=0$) rupture with constant friction under the approximation that dilation occurs suddenly at the frictional rupture tip, and that no flow occurs outside the growing rupture. This corresponds to the case of an infinitely large permeability increase due to shear-induced dilation. 
Under those assumptions for a constant injection rate, the rupture radius also evolves in a self-similar manner following the diffusion lengthscale $
    R(t) = \lambda(\mathcal{T},\epsilon) \sqrt{4\alpha t}
$ 
but the proportionality factor $\lambda$ now depends on both the fault injection stress parameter $\mathcal{T}$ (estimated with the final permeability of the fault after dilation in eq.~\eqref{eq:T_definition}, similar in the expression of the final fault hydraulic diffusivity $\alpha$), and a dimensionless dilation coefficient 
$\epsilon = \dfrac{\tilde{\delta}^p}{w_o S \Delta p_* }$, where $\tilde{\delta}^p$ is the maximum dilatant opening. It is equal to $\psi_p d_c/2$ when using a linearly decreasing dilation coefficient in the constitutive model described in subsection \ref{subsec:Interface_plasticity}.  
When using such a constitutive model, the shear-induced dilation evolve from 0 at the rupture to its maximum value $\tilde{\delta}^p$ over a finite distance - the dilation is not occurring suddenly but gradually as slip accumulates toward the critical value $d_c$.
    As the friction coefficient remain constant, it is possible to estimate the time-scale $t_\psi$ at which the slip at the injection point (center of the rupture) reaches     $d_c$  using the slip solution for a constant friction / constant permeability solution \cite{SaLe22,Vies24}:
        \begin{equation}
        t_\psi = \frac{1}{ 4 \alpha_o} \left(\frac{d_c G}{f \Delta p_*}\right)^2 
    \end{equation}
   where $\alpha_o$ denotes the initial hydraulic diffusivity of the fault (estimated with the fault initial permeability prior dilation), and $\Delta p_*$ is the characteristic injection pressure (estimated with the fault final permeability). 
    We obtain the corresponding lengthscale $L_\psi =\sqrt{4 \alpha_o t_\psi}$. We should numerically recover the sudden dilatant solution for time (/ rupture ) greater than $t_\psi$ (/$L_\psi$).

We simulate (in 3D) the injection at constant rate in a dilatant interface with constant friction. We set $\nu=0$ to obtain a circular rupture consistent with the analytical solution.
Contrary to the approximation of the analytical solution, we set an initial permeability value and use the shear-zone flow model to simulate the increase of permeability with shear-induced dilation. We set the material parameters, in-situ stress and injection rate to achieve a 50 fold hydraulic transmissibility increase and the following values for the dimensionless fault injection stress and dimensionless dilatant strain:
\begin{equation}
    \mathcal{T}=1.3333 \qquad \epsilon =3.09     \qquad \frac{kw|_{final}}{k_o w_o} =50
\end{equation}
The full set of parameters are listed in Supplemental materials.  It results in $t_\psi=4.8\,10^5$ seconds, and $L_\psi =99 $m. We perform a simulation over a large time and propagation distance, discretizing the fault plane over $100\times L_\psi$ with a total of 20k elements.

\begin{figure}
    \centering
    \includegraphics[width=\linewidth]{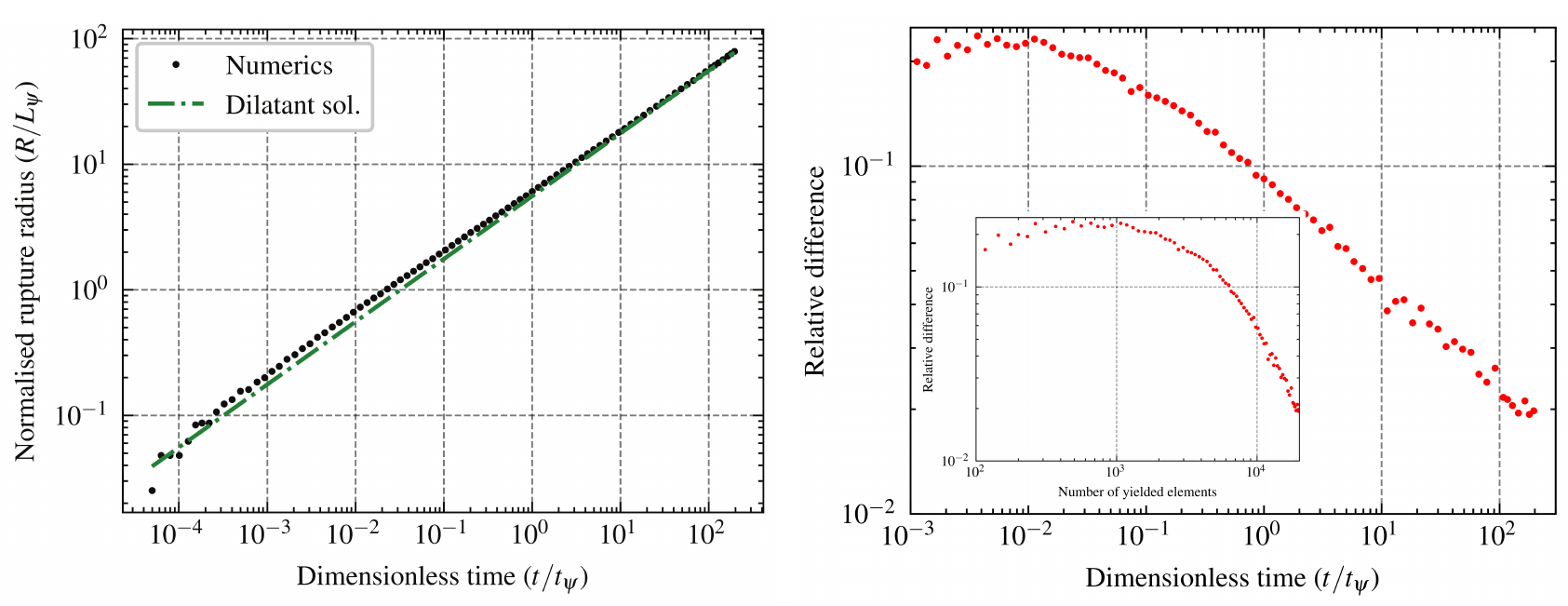}    
    \caption{Evolution of the rupture radius with time over 6 decades of time for a constant friction  dilatant fluid-driven rupture (left). After an initial transient (for $t<t_\psi$), the numerical solution which accounts for linearly decreasing dilatancy with slip tends to the sudden dilation solution \citep{Dunh24} at times larger than the characteristic time $t_\psi$ as expected. The relative difference between the numerical results and sudden dilation solution (right) drops below 4 percent for $t>10 t_\psi$. See Fig.~\ref{fig:Profiles_dilatant_example} for fluid pressure and displacement discontinuity profiles at selected times.
    }
    \label{fig:R_t_dilatant_example}
\end{figure}

\begin{figure}
    \centering
    \includegraphics[width=\linewidth]{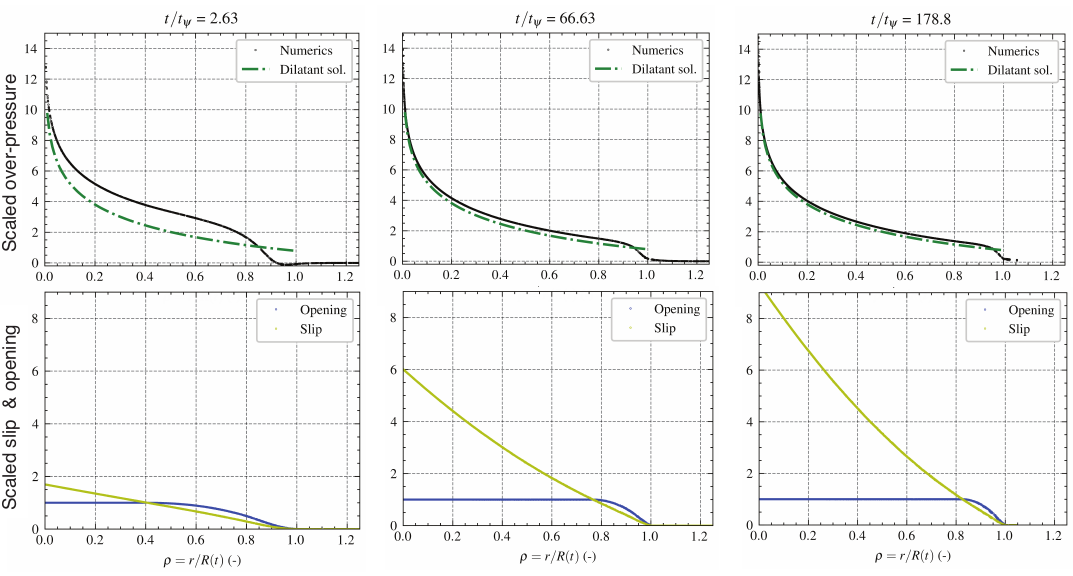}
    \caption{Scaled profiles as function of dimensionless radial distance at different dimensionless times.
    Top row: scaled over-pressure profile $\Delta p/(4\pi \Delta p_*)$ (numerical and sudden dilatant solution); bottom row: scaled slip $\delta/d_c$ and opening $\delta_n^p/\tilde{\delta}^p$ profiles.}
    \label{fig:Profiles_dilatant_example}
\end{figure}

The evolution of the rupture radius of this circular rupture is presented in Fig.~\ref{fig:R_t_dilatant_example} over 6 decades of time, and 3 decades of propagation distance. After an initial transient associated with the evolution of dilatancy towards its final in the near tip region, the numerical results converge toward the sudden dilatant solution \citep{Dunh24} for $t/t_\psi >1$. The relative difference between the numerical results and sudden dilatant solution continuously decrease, and falls below 3 percent for $t>10 t_\psi$. Comparing the evolution of such relative difference with the number of elements within the rupture area (see inset in the left Fig.~\ref{fig:R_t_dilatant_example}), with the one presented for the non-dilatant case in \ref{fig:lam_plot}
indicate that the decrease of the difference is clearly associated with the evolution of dilatancy along the rupture (and not the lack of resolution associated with the discretization).
This physical effect can be better grasped on Fig.\ref{fig:Profiles_dilatant_example}, where the over-pressure $p-p_o$ profiles and the scaled slip $\delta/d_c$ and opening $\delta_n^p/\tilde{\delta}^p$ are plotted for different times 
as function of radial distance from the injection point.
As time progress, the evolution of dilatant opening with slip towards its maximum value becomes more confined toward the rupture tip. As a result the fluid over-pressure profiles obtained numerically converge toward the sudden dilatant solution.

    This  example verify the accuracy of the numerical solver in the case where non-linear hydro-mechanical effects associated with dilation and permeability changes are activated. 

\subsection{Hydraulic fracture propagation along a existing discontinuity}

Thus far, in the previous example, the injection over-pressure always remained below the initial effective normal stress acting on the interface - such that only frictional shear failure was activated. 
We verify here the tensile hydraulic fracture mode propagation by letting the injection pressure going above the normal in-situ traction. The activation of hydraulic fracture opening and the associated very non-linear permeability changes associated with the cubic law render the system of equations very stiff \citep{AdSi07,PeDe08}. 
We model fluid injection at a constant volumetric rate $Q_o$ from a point source in a 3D planar discontinuity having a finite cohesion that linearly weaken with accumulated opening - with therefore a finite mode I fracture energy (toughness $K_{Ic}=\sqrt{E^\prime \sigma_t w_c/2}$). 
No initial shear traction are set on the plane such that upon fluid injection above the normal effective traction, a solely tensile hydraulic fracture propagates in a penny-shaped geometry.

\begin{figure}
    \centering
    \includegraphics[width=0.8\linewidth]{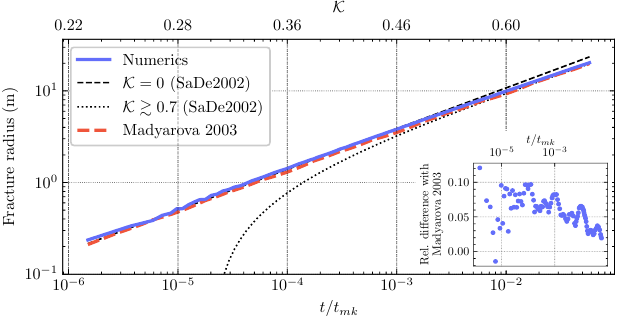}
    \caption{Evolution of the fracture radius of a penny-shaped hydraulic fracture, spanning part of the transition from the viscosity-dominated regime to the toughness-dominated regime. $\mathcal K = 0$ and $\mathcal K \gtrsim 0.7$ are respectively the viscosity regime solution and of the first order viscosity correction solution to the toughness regime, as derived by \cite{SaDe02}. We also show the high fidelity numerical results obtained by \citep{Mady03} and the relative difference between these results and ours.}
    \label{fig:mtok_radius}
\end{figure}

\begin{figure}
    \centering
    \includegraphics[width=\linewidth]{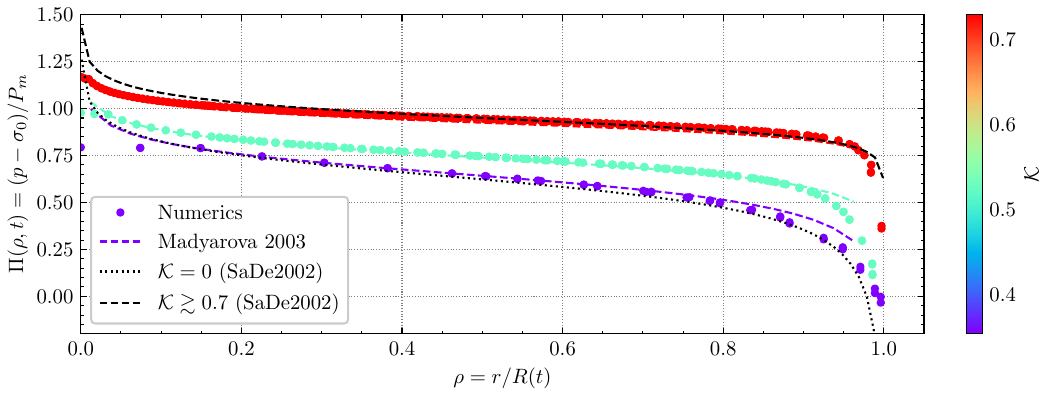}
    \vspace{0.5em}
    \includegraphics[width=\linewidth]{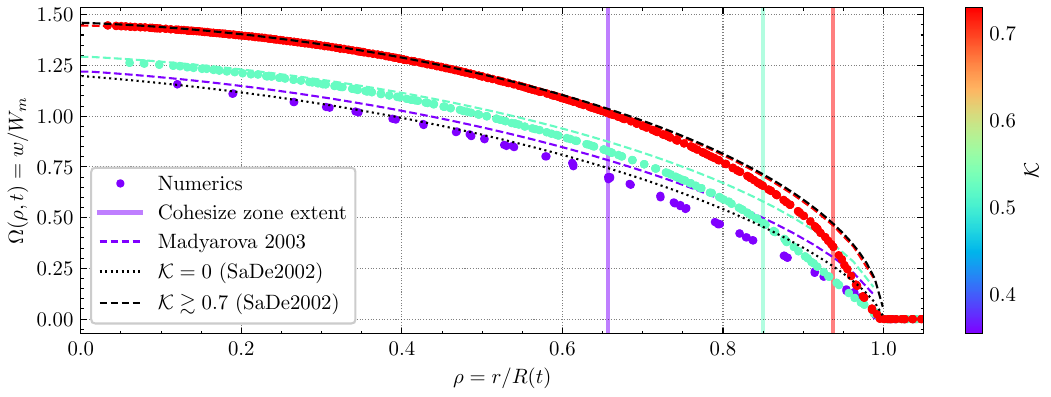}
    \caption{Profiles along the dimensionless radius $\rho=r/R$ of dimensionless pressure $\Pi$ and opening $\Omega$ for three values of dimensionless toughness $\mathcal K = (t / t_{mk})^{1/9}$. We observe the transition from the viscosity solution, denoted $\mathcal K=0$, to the toughness solution with  a first order correction for viscosity, denoted $\mathcal K \gtrsim 0.7$. We also show the numerical results obtained by \citep{Mady03} for reference. 
    The position of the end of the cohesive zone used in our numerical results is also reported. Our numerical results of course differ from the HF solutions based on LEFM  within and close to the process zone associated with the cohesive zone model used.}
    \label{fig:mtok_profiles}
\end{figure}

\begin{figure}
    \centering
    \includegraphics[width=\linewidth]{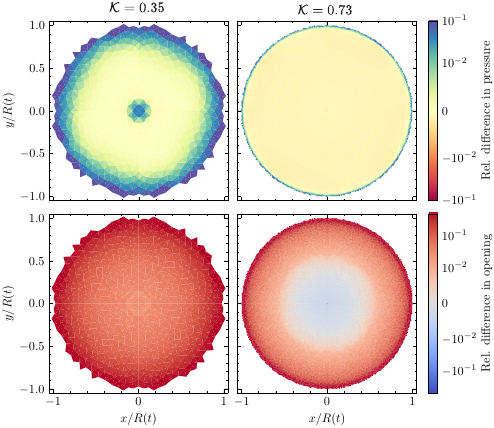}
    \caption{Relative difference in pressure and opening between our results and the high fidelity numerical results obtained by \citep{Mady03}, for two values of dimensionless toughness.}
    \label{fig:mtok_2d_rel_error}
\end{figure}

We can thus compare our numerical results with the penny-shaped hydraulic fracture (HF) growth solutions for the case of an impermeable medium \citep{SaDe02}. 
Notably, it is well known, that the HF transitions from a regime where viscous dissipation dominates (viscosity-storage dominated M-regime) toward a regime where fracture energy dissipation dominates (toughness-storage dominated K-regime) at large time. The scalings can be recovered directly from energy arguments \citep{PeMo24}. 
Self-similar solutions exists in both regimes (and for first order perturbation in dimensionless toughness/viscosity), and high resolution numerical results \citep{Mady03} and approximate solutions \citep{Dont16} exist spanning the regimes transition. 
The transition occurs over the characteristic time-scale $t_{mk}=\dfrac{E^{\prime 13/2} (12\mu)^{5/2} Q_o^{3/2}}{K_{Ic}^9}$.
More specifically, the transition between the viscosity and toughness regime is better grasp via the time-dependent dimensionless toughness $\mathcal{K}=(t/t_{mk})^{1/9}$, with the viscosity regime valid for $\mathcal{K}< 0.31$, the toughness regime for $\mathcal{K}>1.1$ (and the first-order viscosity correction valid for $\mathcal{K}\gtrsim 0.7$). We refer to \cite{SaDe02} for details of the solution. 
The material, in-situ stress and injection parameters used are listed in Supplemental materials (as well as the mesh and numerical simulation tolerance). We perform a simulation spanning the viscosity regime, and the transition toward the toughness regime. This is particularly important to demonstrate -in a single simulation- the capability of the solver to capture the different self-similar behavior of these two regimes - and accuracy over several time and length-scales. Figure \ref{fig:mtok_radius} shows the evolution of the fracture radius with dimensionless time $t/t_{mk}$ (or equivalently dimensionless toughness $\mathcal K$) along with the zero toughness asymptote of the viscosity regime ($\mathcal K = 0$) and the first order viscosity correction of the large toughness asymptote associated with the toughness  regime ($\mathcal K \gtrsim 0.7$). We also show the results obtained by \citep{Mady03} as reference along with the relative difference between these results and ours. Figure \ref{fig:mtok_profiles} show profiles of dimensionless pressure and opening at three different dimensionless toughness values, here as well along with the self-similar asymptotes of the viscosity regime and the toughness regime with first order correction for viscosity. In both cases we scale the results using the time-dependent scalings associated with the viscosity regime $P_m$ and $W_m$:
\[
    P_{m}=\left(\frac{12\mu E^{\prime 2}}{t}\right)^{1/3} \qquad W_{m} =\left(\frac{(12\mu)^{2} Q_0^3 t}{E^{\prime 2}}\right)^{1/3}
\]
We observe that the obtained results are in good agreement with the expected analytical solutions for the limiting regimes as well as the numerical results of \citep{Mady03}. For the fracture opening profiles, we note that our numerical opening deviates from the LEFM solution within and close to the process zone associated with the cohesive zone model, which is expected (see 
\cite{LePe13} for further discussion).
Figure \ref{fig:mtok_2d_rel_error} displays the spatial repartition of the relative error in pressure and opening with respect to the results of \citep{Mady03} for two values of dimensionless toughness.

\subsection{Computational Performance Insights \label{subsec:computational_insights} 
}

We showcase the computational performance of the solver by running two set of simulations, varying the number of degrees of freedom. The first set of simulations model the propagation of a frictional rupture, considering both the permeability and the friction coefficient to be constant. The second set of simulation model the propagation of a viscosity-dominated hydraulic fracture. Both simulations are ran over the same three 3D planar disk domains. The associated meshes have the same resolution at the injection point and the same mesh size gradient. Their size have been chosen so to obtain number of mechanical degrees of freedom of about 8'000, 32'000 and 128'000. We run these simulations using the two possible approximations of ${\bf A}_{11}^{-1}$ as discussed in Section \ref{section_sol_jac_system}. The first one is to take $\tilde {\bf A}_{11} = diag({\bf A}_{11})$, referenced to as ``Jacobi'', the second one is to take $\tilde {\bf A}_{11}$ as a sparse subset of the entries of ${\bf A}_{11}$ and to then compute its ILU factorization, referenced to as ``ILU(0)''.
For each mesh size, we report metrics at three snapshots during the simulation corresponding to 20\%, 40\%, and 60\% of elements in a yielded state. The yielded fraction has a first-order effect on the conditioning of the mechanical sub-system ${\bf A_{11}}$: as more elements yield, ${\bf A}_{11}$ becomes less diagonally dominant (in relation to the contributions of the consistent tangent operator). This lead to an increase in the number of iterations required to solve  ${\bf A}_{11}$ but also decrease the quality of the $\tilde{{\bf A}}_{11}$ approximate used in the Schur complement approximation, resulting in an increase of the number of iterations required to solve on the block tangent system.
Four metrics are reported in Tables \ref{tab:perf_slip} and \ref{tab:perf} as medians over ten time-steps: total compute time per time-step, Newton iterations per time-step, iterative solver iterations on the full tangent block system (``Jacobian iterations''), and iterative solver iterations on the mechanical sub-system. Here in both case we rely on the BICGSTAB iterative solver.  

The first observation is that the hydraulic fracture simulations, where the hydraulic conductivity grows that the cube of the mechanical opening, require much more iterations of the iterative solver on the tangent system compare to the frictional ruptures simulations that consider a constant permeability. Newton convergence is robust across all configurations, requiring four to five iterations regardless of mesh size, preconditioner, or yielded fraction. Using the ILU(0) approximate of ${\bf A}_{11}^{-1}$, rather than the Jacobi, yields a substantial improvement of the quality of the overall preconditioning of the tangent block system in the hydraulic fracture simulations but yields am even better improvement in the preconditioning of ${\bf A}_{11}$ itself. This is especially true for the frictional rupture simulations where the consistent tangeant operator blocks of yielded elements have non diagonal terms that are not captured by the simpler Jacobi inverse approximation.
The simulation were run on an HPC compute node with 16 threads of an AMD EPYC 9334 @ 2.7 GHz, associated with an NVIDIA H100 GPU that provides acceleration of the BEM matrix-vector multiplication and of the flow FEM matrices assembly, these two operations make up the most part of the computational cost of the overall solver.


\begin{table}[]
\centering
\setlength{\tabcolsep}{3pt}
\renewcommand{\arraystretch}{1.3}
\begin{tabular}{ll | SSS | SSS | SSS}
\toprule
& & \multicolumn{3}{c}{\textbf{2.5k elements}} & \multicolumn{3}{c}{\textbf{10k elements}} & \multicolumn{3}{c}{\textbf{40k elements}} \\
\cmidrule(lr){3-5} \cmidrule(lr){6-8} \cmidrule(lr){9-11}
\textbf{Metric} & \textbf{${\bf A}_{11}^{-1}$} app. 
& {20\%} & {40\%} & {60\%} 
& {20\%} & {40\%} & {60\%} 
& {20\%} & {40\%} & {60\%} \\
\midrule

\multirow{2}{*}{\makecell{Compute \\time (s)$^{1}$}}
 & Jacobi & 0.49 & 0.53 & 0.57 & 1.1 & 1.4 & 1.5 & 7.5 & 11 & 15 \\
 & ILU(0) & 0.36 & 0.40 & 0.42 & 0.52 & 0.57 & 0.62 & 2.0 & 2.3 & 2.7 \\
\cmidrule(l){1-11}
 
\multirow{2}{*}{\makecell{Newton \\iterations$^{1}$}}
 & Jacobi & 4.2 & 4.0 & 4.0 & 4.0 & 4.0 & 4.0 & 4.0 & 4.0 & 4.0 \\
 & ILU(0) & 4.2 & 4.0 & 4.0 & 4.0 & 4.0 & 4.0 & 4.0 & 4.0 & 4.0 \\
\cmidrule(l){1-11}

\multirow{2}{*}{\makecell{Jacobian \\iterations$^{2}$}}
 & Jacobi & 1.7 & 1.6 & 1.7 & 1.7 & 1.7 & 1.7 & 1.7 & 1.7 & 1.7 \\
 & ILU(0) & 1.5 & 1.6 & 1.7 & 1.7 & 1.7 & 1.7 & 1.7 & 1.7 & 1.60 \\
\cmidrule(l){1-11}
 
\multirow{2}{*}{\shortstack[l]{${\bf A}_{11}$\ solver\\iterations$^{2}$}}
 & Jacobi & 102 & 111 & 133 & 132 & 174 & 188 & 228 & 347 & 482 \\
 & ILU(0) & 21.7 & 24.0 & 27.4 & 24.9 & 32.2 & 32.9 & 35.1 & 43.6 & 44.4 \\
 
\bottomrule
\end{tabular}
\caption{Frictional rupture with constant permeability and constant friction: solver performance metrics (median over ten time-steps) as function of the number of elements for different fractions of yielded elements (with respect to the total number of elements), corresponding to different times throughout the simulations.
We report these statistics for the two possible choices used to approximate the inverse of the ${\bf A}_{11}$ block of the tangent hydro-mechanical system ($^{1}$~per time step, 
$^{2}$~per linear solve (BICGSTAB)).}
\label{tab:perf_slip}
\end{table}

\begin{table}[]
\centering
\renewcommand{\arraystretch}{1.3}
\begin{tabular}{ll | SSS | SSS | SSS}
\toprule
& & \multicolumn{3}{c}{\textbf{2.5k elements}} & \multicolumn{3}{c}{\textbf{10k elements}} & \multicolumn{3}{c}{\textbf{40k elements}} \\
\cmidrule(lr){3-5} \cmidrule(lr){6-8} \cmidrule(lr){9-11}
\textbf{Metric} & \makecell{\textbf{${\bf A}_{11}^{-1}$} \\ \textbf{approximate}} & {20\%} & {40\%} & {60\%} & {20\%} & {40\%} & {60\%} & {20\%} & {40\%} & {60\%} \\
\midrule

\multirow{2}{*}{\makecell{Compute \\time (s)$^{1}$}}
 & Jacobi & 0.75 & 1.06 & 1.21 & 3.51 & 4.72 & 6.47 & 34.91 & 51.43 & 56.80 \\
 & ILU(0) & 0.41 & 0.70 & 0.72 & 1.87 & 2.54 & 3.88 & 13.93 & 25.95 & 36.05 \\
\cmidrule(l){1-11}
 
\multirow{2}{*}{\makecell{Newton \\iterations$^{1}$}}
 & Jacobi & 3.5 & 4.0 & 4.0 & 4.0 & 4.3 & 5.0 & 5.0 & 4.3 & 4.5 \\
 & ILU(0) & 3.0 & 4.0 & 4.0 & 4.0 & 4.0 & 5.0 & 4.9 & 5.0 & 5.0 \\
\cmidrule(l){1-11}

\multirow{2}{*}{\makecell{Jacobian \\iterations$^{2}$}}
 & Jacobi & 8.5 & 10.0 & 10.0 & 11.5 & 13.0 & 13.0 & 14.4 & 19.0 & 20.0 \\
 & ILU(0) & 5.0 & 6.0 & 7.5 & 7.0 & 8.0 & 9.4 & 9.0 & 11.5 & 15.0 \\
\cmidrule(l){1-11}
 
\multirow{2}{*}{\shortstack[l]{${\bf A}_{11}$\ solver\\iterations$^{2}$}}
 & Jacobi & 27.5 & 34.0 & 39.5 & 41.2 & 49.8 & 54.7 & 61.4 & 71.5 & 79.9 \\
 & ILU(0) & 14.0 & 16.4 & 18.3 & 18.5 & 22.5 & 26.0 & 23.4 & 31.2 & 38.7 \\
 
\bottomrule
\end{tabular}
\caption{Viscosity-dominated hydraulic fracture: solver performance metrics (median over ten time-steps) as function of the number of elements for different fraction of the number of yielded elements (with respect to the total number of elements) corresponding to different times throughout the simulations.
We report these statistics for the two possible choices to approximate the inverse of the ${\bf A}_{11}$ block of the tangent hydro-mechanical system ($^{1}$~per time step, 
$^{2}$~per linear solve (BICGSTAB)).}
\label{tab:perf}
\end{table}

\section{Examples of fluid-driven ruptures involving several discontinuities}

\subsection{Injection into a set of intersecting fractures}

We first investigate fluid injection into three intersecting circular fractures. This configuration is introduced as a minimal extension of the single-fracture problem, while still allowing for hydraulic communication between fractures and for mechanical stress transfer associated with slip on connected discontinuities.

The network consists of one central fracture, into which fluid is injected at its center, and two secondary fractures intersecting it on opposite sides. In the following, we refer to these secondary fractures as the left and right fractures. The central fracture has a radius of $90$ m and dips at $65^\circ$, while the left and right fractures have smaller radii, equal to $75$ m and $60$ m, respectively. Both secondary fractures intersect the central fracture, thereby providing two connected hydraulic pathways away from the injection point.

\begin{figure}[htbp]
    \centering
    \begin{subfigure}[b]{0.48\linewidth}
        \centering
        \includegraphics[width=\linewidth]{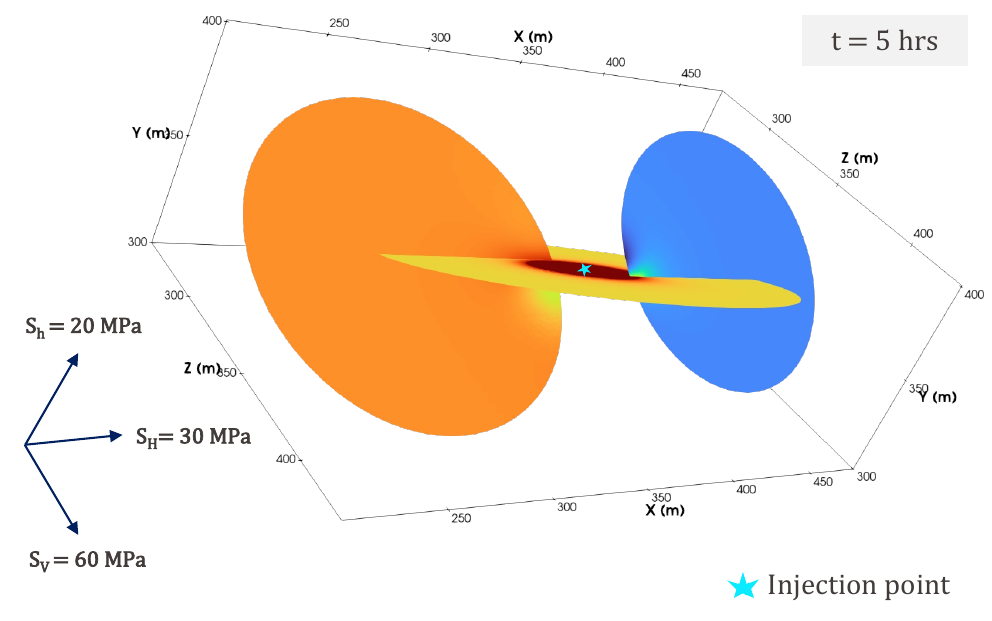}
        \label{fig:yf_2hrs}
    \end{subfigure}
    \hfill
    \begin{subfigure}[b]{0.47\linewidth}
        \centering
        \includegraphics[width=\linewidth]{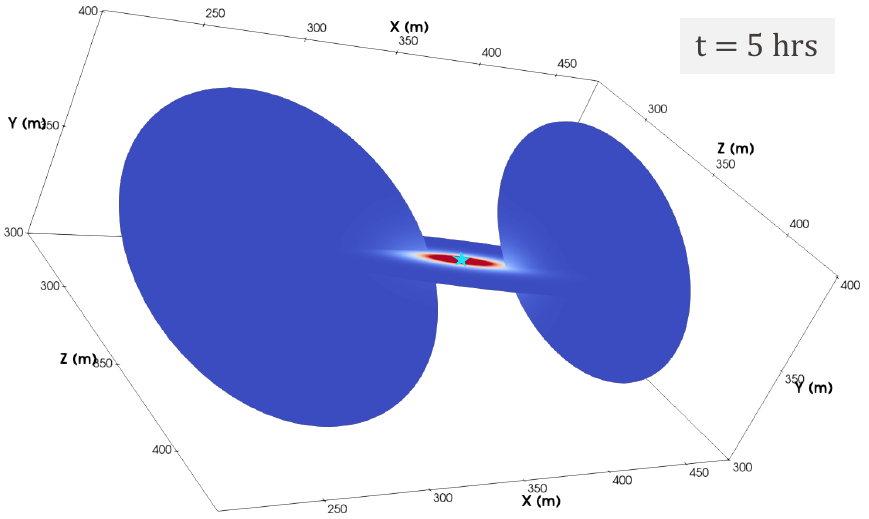}
        
        \label{fig:pressure_2hrs}
    \end{subfigure}

    \begin{subfigure}[b]{0.48\linewidth}
        \centering
        \includegraphics[width=\linewidth]{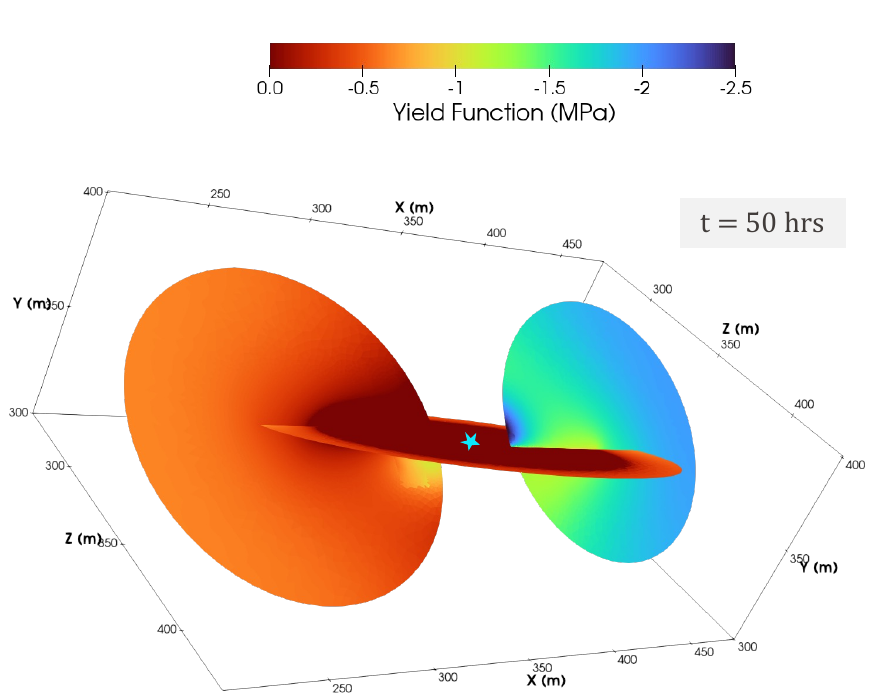}
        
        \label{fig:yf_20hrs}
    \end{subfigure}
    \hfill
    \begin{subfigure}[b]{0.45\linewidth}
        \centering
        \includegraphics[width=\linewidth]{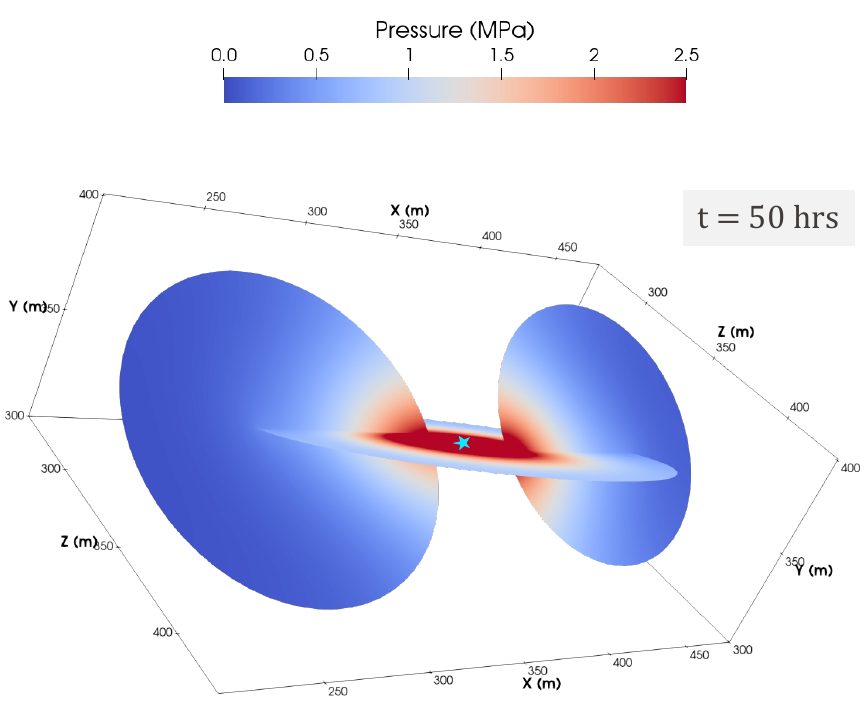}
        
        \label{fig:pressure_20hrs}
    \end{subfigure}

    \caption{Evolution of the yield function and fluid pressure in the three-fracture configuration at \(t=5\) h and \(t=50\) h. The left column shows the spatial distribution of the yield function on the fracture surfaces, with values approaching zero indicating regions closer to frictional yielding, while the right column shows the corresponding pressure field.}
    \label{fig:yf_pressure_threefrac}
\end{figure}

The imposed in-situ stress state is prescribed from the three principal stresses $S_V=60$ MPa, $S_H=30$ MPa, and $S_h=20$ MPa, whose orientations are shown in Fig.~\ref{fig:yf_pressure_threefrac}. Although the far-field stress tensor is uniform, the different orientations of the three fractures result in different local initial tractions. We quantify the initial level of shear criticality of each fracture using the pre-stress ratio $\mathcal{S}=\tau_0/(f\sigma'_0)$. For the present configuration, the central, left, and right fractures have comparable initial criticalities, with $\mathcal{S}=0.94$, $0.96$, and $0.90$, respectively.

The fracture surfaces are discretized with 89'223 triangular elements and 44'685 nodes, resulting in a total of 312'354 hydro-mechanical degrees of freedom. The mesh is refined near the fracture intersections in order to better resolve hydraulic communication between fractures and the associated localized mechanical response. We perform the simulation for a constant injection rate, assuming a constant friction coefficient, constant permeability, and zero dilatancy. The complete set of parameters and mesh characteristics used for this simulation are reported in the Supplemental Material.

As fluid is injected in fracture 1, the pore-pressure perturbation first diffuses along the injection fracture and a slipping patch nucleates around the injection point. The subsequent growth of this patch modifies the tractions on the two intersecting fractures. This configuration activates both hydraulic communication through the intersections and elastic stress transfer between slipping discontinuities.

The evolution of the yield function at $t=5$ h and $t=50$ h is displayed in Fig.~\ref{fig:yf_pressure_threefrac}. At early time ($t=5$ h), the yielded region remains mainly localized on the central fracture around the injection point. However, the stress perturbation induced by slip on the central fracture is already visible on the two secondary fractures. In particular, the upper part of the right fracture and the lower part of the left fracture experience a decrease of the yield function. In these regions, elastic stress transfer therefore acts in a stabilizing way, moving the fractures locally farther from frictional failure. At later time ($t=50$ h), the slipping patch has further propagated along the central fracture and the mechanical interaction with the intersecting fractures becomes more pronounced. Pressure diffusion tends to bring the secondary fractures closer to frictional yielding, while the zones previously stabilized by elastic stress transfer shield part of these fractures from reactivation. As a result, the yielded region does not propagate symmetrically on the two sides of the network. Slip preferentially develops toward the upper part of the left fracture, while the lower part of the right fracture also moves closer to yielding. The opposite branches remain comparatively stabilized.

This example illustrates that the rupture path in a connected fracture network is not controlled by pore-pressure diffusion alone. Elastic stress transfer associated with slip on one fracture can either promote or inhibit slip on neighboring fractures, depending on their relative position and orientation. 

\subsection{A height confined hydraulic fracture intersecting a fault}

\begin{figure}
    \centering
    \includegraphics[width=\linewidth]{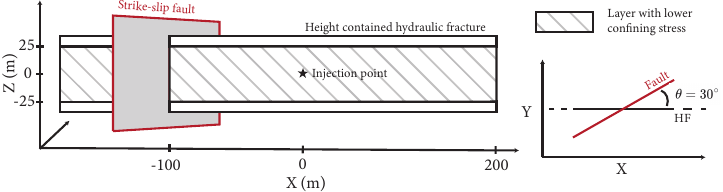}
    \caption{Schematic view of the simulation setup for modeling the intersection of a vertical height contained hydraulic fracture with a vertical strike slip fault.}
    \label{fig:hf_fault_setup}
\end{figure}

The last example aims to demonstrate how this mix mode solver enables us to tackle the numerically challenging problem of an hydraulic fracture intersecting a fault in 3D. We model the intersection of a vertical height contained hydraulic fracture with a also vertical strike slip fault as depicted in Figure \ref{fig:hf_fault_setup}. The hydraulic fracture is oriented in the $xz$ plane while the strike fault makes an angle of $30^\circ$ with the $x$ direction. We suppose a strike slip stress regime with the vertical stress and pore pressure being respectively lithostatic and hydrostatic, with a rock density of $2700$ kg/m$^3$ and at a depth of $2000$ m. We use different values of horizontal stresses depending inside and outside the central layer. The minimum horizontal stresses are set to values that ensure the height confinement of the hydraulic fracture within the central layer (the values are reported Supplemental Materials). The maximum horizontal stresses in the different layers are obtained such that everywhere on the fault we have $\mathcal S = \tau / (f\sigma_n^{\prime}) = 0.99$ ($\tau$ the shear stress and $\sigma_n^{\prime}$ the normal effective stress). We set the flow constitutive law as the cubic law for the elements in the hydraulic fracture plane and as the shear zone flow in the fault plane. All material and in-situ stress parameters are reported in the Supplemental Materials.

Figure \ref{fig:hf_fault_3d} displays the distribution of opening (on the HF plane), and slip (on the fault plane) at two instant: one before the HF has intersected the fault and one at the end of the simulation where the fault slip reaches the boundary of the fault domain. We observe that for our set of parameters the HF is arrested upon fault intersection. We also observe some slip onto the fault by stress transfer before the intersection with the HF occur, while most of the slip at late time is due to the diffusion of the over-pressure onto the fault. Figure \ref{fig:hf_fault_profiles} show the same data for the same two time steps by means of profiles along the mean horizontal line on both the fault plane and the HF plane.

\begin{figure}
    \centering
    \includegraphics[width=0.8\linewidth]{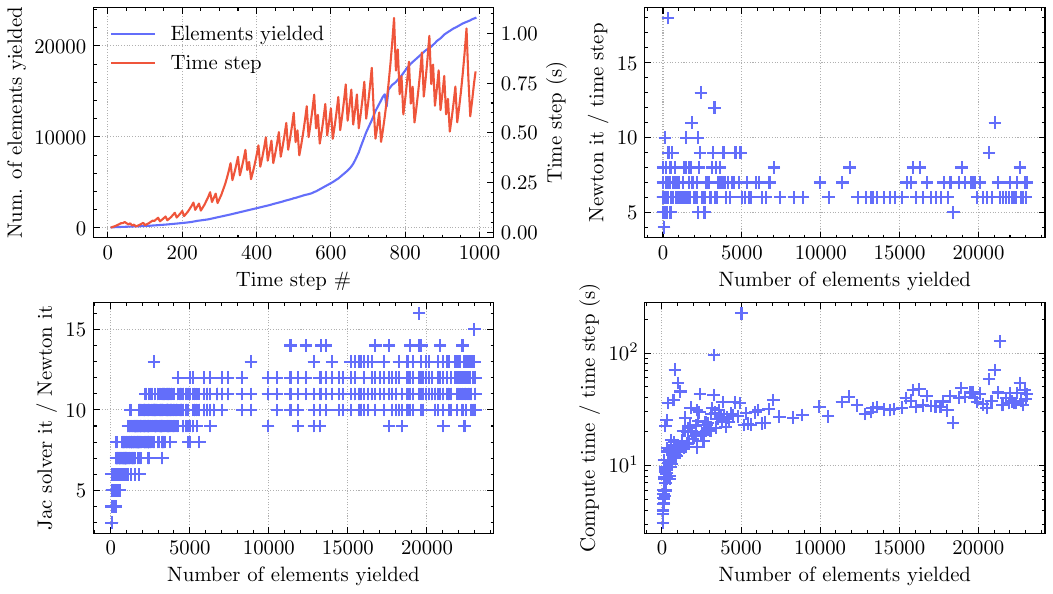}
    \caption{Numerical solver statistics for the HF / fault interaction simulation: i) time-step size evolutions, and the number of element yielded, ii) Newton iterations per time-step, iii) Number of iteration of the Jacobian solver for a Newton iteration, iv) compute time of a time-steps as function of the number of yielded elements.}
    \label{fig:hf_fault_statistics}
\end{figure}

The mesh contains  40'646 triangular surface elements for about 20'549 nodes which makes a total of 142'487 hydromechanical degrees of freedom. It is important to keep in mind that a solver based on a volumetric discretization for mechanical deformation (for example finite element) would require more than 10 millions of degrees of freedom to achieve the same resolution. 
The simulation runs in about 8 hours with the BEM matrix-vector multiplication as well as the flow FEM matrices assembly benefiting from GPU acceleration. We show in Figure \ref{fig:hf_fault_statistics} some statistics of the numerical solver throughout the simulation. We observe reasonable number of iterations for the Newton solver and the block Jacobian linear solver. Both linear iterative solvers require slightly more iterations as more elements yield as the ${\bf A}_{11}$ block becomes less diagonally dominant, as discussed in \ref{subsec:computational_insights}.

 \begin{figure}
    \centering
    \includegraphics[width=\linewidth]{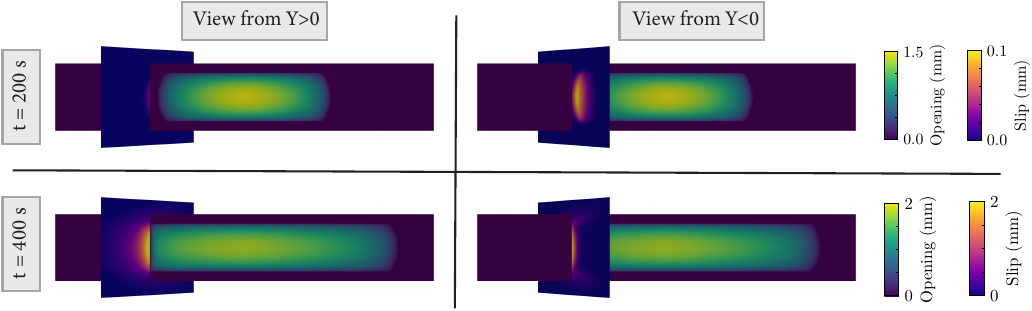}
    \caption{Numerical results of the height contained HF interacting with a strike slip fault. Opening on the HF plane along with the slip on the fault plane for two times: before (top) and after (bottom) the HF intersect the fault.}
    \label{fig:hf_fault_3d}
\end{figure}

 \begin{figure}
    \centering
    \includegraphics[width=\linewidth]{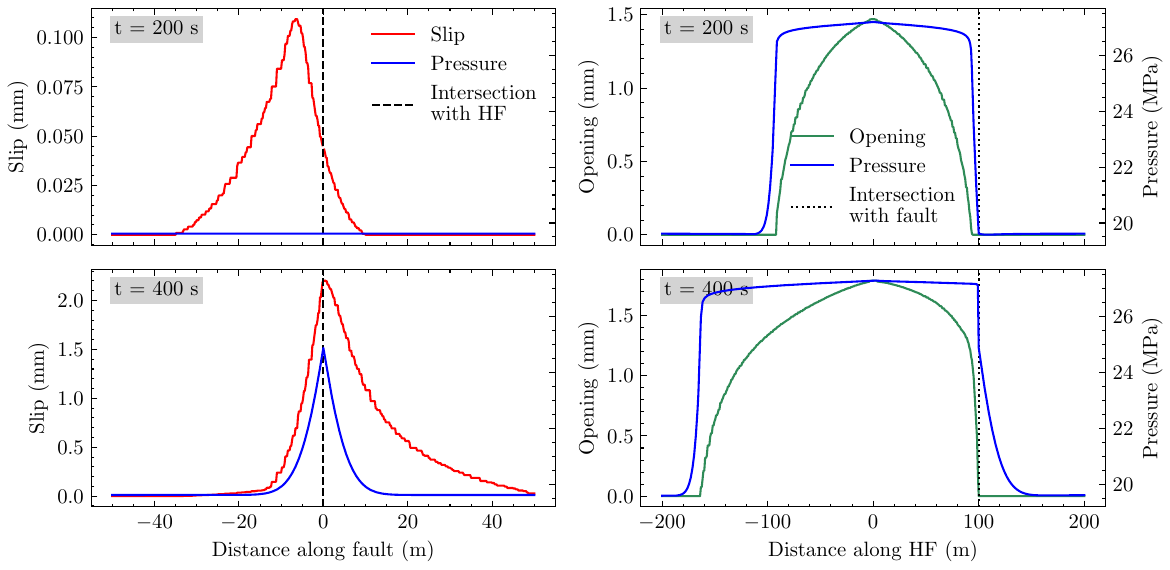}
    \caption{Profiles of slip, opening, and pressure taken along the intersections of the fault and HF planes with the horizontal plane at $z=0$ (depth of injection). For the same two time steps represented in Figure \ref{fig:hf_fault_3d}.}
    \label{fig:hf_fault_profiles}
\end{figure}

 \begin{figure}
    \centering
    \includegraphics[width=0.8\linewidth]{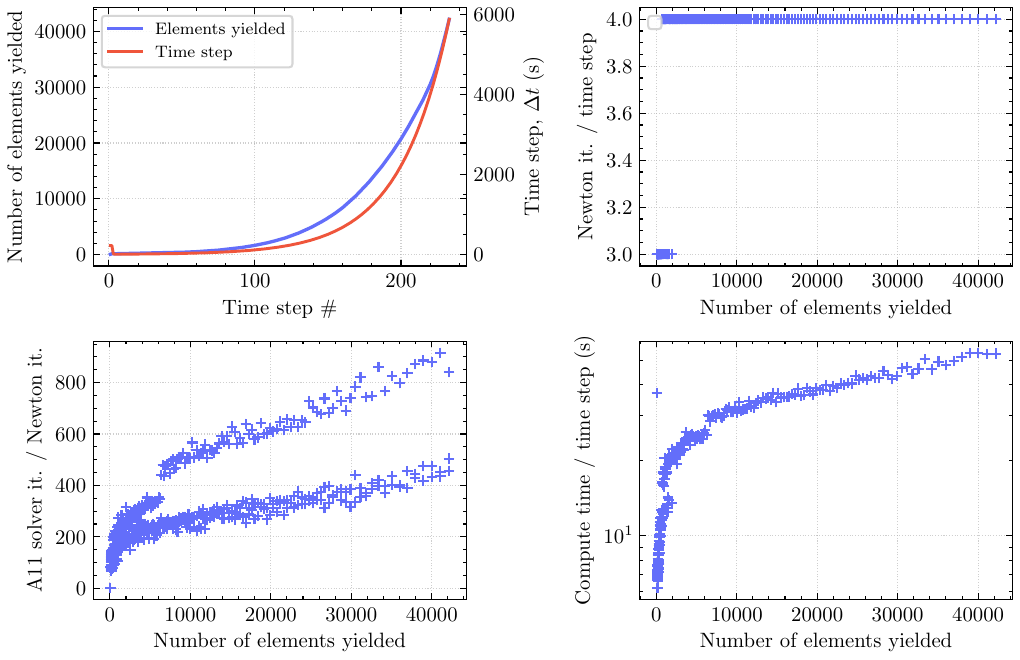}
    \caption{Numerical solver statistics for the three fracture simulation: i) time-step size evolutions, and the number of element yielded, ii) Newton iterations per time-step, iii) Number of iteration of the A11 Krylov solver for a Newton iteration, iv) compute time of a time-steps as function of the number of yielded elements.}
    \label{fig:three_frac_statistics}
\end{figure}

\section{Conclusions and perspectives}

Robust and accurate simulations of fluid-driven fractures growth over multiple length and time scales remains challenging. We have presented a implicit fully-coupled solver based on a boundary element discretization of the quasi-static balance of momentum and a finite element discretization of the flow operator. 
Rupture along existing (meshed) discontinuities are modeled via an elasto-plastic (cohesive zone) like constitutive formulation accounting for tensile and frictional weakening with dilation. 
The non-linear variation of interface permeability as function of shear-induced dilation and mechanical opening render the hydro-mechanical problem extremely stiff. Proper pre-conditioning of the hydro-mechanical tangent system is even more critical when such non-linearities between permeability and mechanical deformation occur. Simple staggering strategies fail.

The series of verification examples presented here to quantify the accuracy and robustness of the proposed solver is universal. It should help in further developing robust numerical algorithms for this class of fracture problems. 
We do believe that other spatial discretizations for example using domain method such as finite element or finite volume methods for the solution of the mechanical equations will also benefit from the use of these verification tests. 
We anticipate that similar accuracy can be reach likely at the expense of finer discretizations, as finite element typically requires more resolution to achieve the same accuracy than boundary element for fracture mechanics problem. 

The mix-mode cohesive zone like model used has the advantage of simplicity. Additional physics (such as rate and state friction, or more intricate dilation/compaction law via modified Cam-Clay models) can be easily implemented in such a framework - pending the use of the corresponding consistent tangent operator. 
Different fracture energies in tension and shear can be accounted for, and both fluid-driven frictional weakening  and tensile hydraulic fracture solutions were reproduced adequately. 
However, sufficiently fine resolution must be used to properly resolve the fracture process zone (and the associated energy dissipation). 
As such, the proposed algorithm is computationally less efficient than the implicit level set algorithm \citep{PeDe08} specialized for hydraulic fracture growth (and operate using linear elastic fracture mechanics). 

Although we have restricted to the case of an impermeable rock matrix, the formulation can be extended to account for matrix flow via a domain based method (in a multi-dimensional context). This will result in a larger size of the matrix blocks associated with fluid flow in the tangent system. Proper block preconditioning strategy would have to be tested to ensure robustness and efficiency. 
More importantly, semi-analytical propagation solutions accounting for matrix fluid flow (and thermal effects) in the context of frictional ruptures (and to a lesser extent for hydraulic fracture) must be further developed. The availability of rupture propagation solutions for simple fracture geometry is essential to not only verify solver implementation but also further advance numerical schemes for the solution of multiphysics fracture growth problem. 

\paragraph{Declaration of generative AI and AI-assisted technologies in the writing process\\}

During the preparation of this work the authors used Sonnet 4.6 to improve language and readability of part of the text, notably the abstract. After using this tool/service, the authors reviewed and edited the content as needed and take full responsibility for the content of the publication.

\paragraph{CRediT author statement\\}  
\textbf{B. Lecampion}: Conceptualization, Methodology, Software, Validation, Formal analysis, Writing - Original Draft, Resources, Supervision, Project administration, Funding acquisition.
\textbf{S. Brisson}: Methodology, Software, Validation, Formal analysis,
Writing - Original Draft, Visualization
\textbf{A. Sarma}: Methodology, Software, Validation, Formal analysis,
Writing - Original Draft, Visualization
\textbf{A. Gupta}:  Software, Validation, Formal analysis
\textbf{A. S\'aez}: Methodology, Validation, Formal analysis
\textbf{R. Fakhretdinova}: Validation, Formal analysis

\paragraph{Declaration of competing interest\\} 
The authors declare that they have no known competing financial interests or personal relationships that could have appeared to influence the work reported in this paper.

\paragraph{Acknowledgments \\}
The results were partly obtained within the EMOD project (Engineering model for hydraulic stimulation). The EMOD project benefits from a grant (research contract no. SI/502081-01) and an exploration subsidy (contract no. MF-021-GEO-ERK) of the Swiss federal office of energy for the EGS geothermal project in Haute-Sorne, canton of Jura, which is gratefully acknowledged. A.S., R.F. and A.S. were partially funded by the Federal Commission for Scholarships for Foreign Students via the Swiss Government Excellence Scholarship.

\bibliography{./BibDB-Complete}

\end{document}